\documentclass[smallextended]{svjour3}       
\smartqed  									 

\usepackage{graphicx}
\usepackage{amsmath}
\usepackage{amssymb}   
\usepackage{textcmds}  
\usepackage{color}     
\usepackage{gensymb}   
\usepackage[]{hyperref}
\usepackage{upgreek}

\usepackage{natbib}
\bibpunct{(}{)}{;}{a}{}{,}
\newcommand{\farcs}{\mbox{\ensuremath{.\!\!^{\prime\prime}}}}

\newcommand*\aap{A\&A}

\newcommand*\apj{ApJ}
\newcommand*\apjl{ApJ}

\newcommand*\apjs{ApJS}

\newcommand*\mnras{MNRAS}

\newcommand*\nat{Nature}

\newcommand*\pasp{PASP}

\newcommand*\procspie{Proc SPIE}

\begin{document}

\title{The path towards high-contrast imaging with the VLTI: the Hi-5 project}

\titlerunning{High-contrast interferometry with the VLTI}        
\author{D.~Defr\`ere$^1$, O.~Absil$^1$, J.-P.~Berger$^2$, T.~Boulet$^1$, W.C.~Danchi$^3$, S.~Ertel$^4$, A.~Gallenne$^5$, F.~H\'enault$^{2}$, P.~Hinz$^{4}$, E.~Huby$^{6}$, M.~Ireland$^{7}$, S.~Kraus$^{8}$, L.~Labadie$^{9}$, J.-B.~Le Bouquin$^{2}$, G.~Martin$^{2}$, A.~Matter$^{10}$, A.~M\'erand$^{11}$, B.~Mennesson$^{12}$, S.~Minardi$^{13,17}$, J.~Monnier$^{14}$, B.~Norris$^{15}$, G.~Orban de Xivry$^{1}$, E.~Pedretti$^{16}$, J.-U.~Pott$^{17}$, M.~Reggiani$^{1}$, E.~Serabyn$^{12}$, J.~Surdej$^1$, K.\,R.\,W.~Tristram$^{5}$, and J.~Woillez$^{11}$.}

\authorrunning{Defr\`ere et al.}

\institute{D.\, Defr\`ere \at
           Tel.: +32-4-3669758\\
           \email{ddefrere@uliege.be}\\           
           \and
           $^1$ Space sciences, Technologies \& Astrophysics Research (STAR) Institute, University of Li\`ege, Li\`ege, Belgium\\
           \and
           $^2$ Univ. Grenoble Alpes, CNRS, IPAG, 38000 Grenoble, France\\
           \and
           $^3$ NASA Goddard Space Flight Center, Exoplanets \& Stellar Astrophysics Laboratory, Greenbelt, USA\\
           \and
           $^4$ Steward Observatory, Department of Astronomy, University of Arizona, Tucson, Arizona, USA\\       
           \and
           $^5$ European Southern Observatory, Alonso de C\'ordova 3107, Vitacura, Santiago de Chile, Chile\\ 
           \and
           $^6$ LESIA, Observatoire de Paris, PSL Research University, 92195 Meudon Cedex, France\\
           \and
           $^7$ Research School of Astronomy and Astrophysics, Australian National University, Canberra, ACT 2611, Australia\\
           \and
           $^8$ School of Physics and Astronomy, University of Exeter, Exeter, United Kingdom\\ 
           \and
           $^9$ I. Physikalisches Institut, Universit\"at zu K\"oln, Z\"ulpicher Str. 77, 50937 Cologne, Germany\\
           \and
           $^{10}$ Laboratoire Lagrange, Universit\'e C\^ote d'Azur, Observatoire de la C\^ote d'Azur, CNRS, Boulevard de l'Observatoire, CS 34229, 06304, Nice, France\\ 
           \and
           $^{11}$ European Southern Observatory, Munich, Germany\\ 
           \and
           $^{12}$ Jet Propulsion Laboratory, California Institute of Technology, Pasadena, CA 91109, USA\\
           \and
           $^{13}$ University of Jena, Jena, Germany\\
           \and
           $^{14}$ University of Michigan, Ann Arbor, United States\\
           \and
           $^{15}$ University of Sydney, Sydney, Australia\\           
           \and
           $^{16}$ innoFSPEC, Leibniz-Institut für Astrophysik Potsdam (AIP) Germany\\ 
           \and
           $^{17}$ Max Planck Institute for Astronomy, Heidelberg, Germany\\
}

\date{Received: date / Accepted: date}
\maketitle

\begin{abstract}
The development of high-contrast capabilities has long been recognized as one of the top priorities for the VLTI. As of today, the VLTI routinely achieves contrasts of a few 10$^{-3}$ in the near-infrared with PIONIER (H band) and GRAVITY (K band). Nulling interferometers in the northern hemisphere and non-redundant aperture masking experiments have, however, demonstrated that contrasts of at least a few 10$^{-4}$ are within reach using specific beam combination and data acquisition techniques. In this paper, we explore the possibility to reach similar or higher contrasts on the VLTI. After reviewing the state-of-the-art in high-contrast infrared interferometry, we discuss key features that made the success of other high-contrast interferometric instruments (e.g., integrated optics, nulling, closure phase, and statistical data reduction) and address possible avenues to improve the contrast of the VLTI by at least one order of magnitude. In particular, we discuss the possibility to use integrated optics, proven in the near-infrared, in the thermal near-infrared (L and M bands, 3-5~$\upmu$m), a sweet spot to image and characterize young extra-solar planetary systems. Finally, we address the science cases of a high-contrast VLTI imaging instrument and focus particularly on exoplanet science (young exoplanets, planet formation, and exozodiacal disks), stellar physics (fundamental parameters and multiplicity), and extragalactic astrophysics (active galactic nuclei and fundamental constants). Synergies and scientific preparation for other potential future instruments such as the Planet Formation Imager are also briefly discussed. 
\end{abstract}

\section{Introduction}
\label{intro}

\begin{figure}[!t]
	\begin{center}
		\includegraphics[height=8.1 cm]{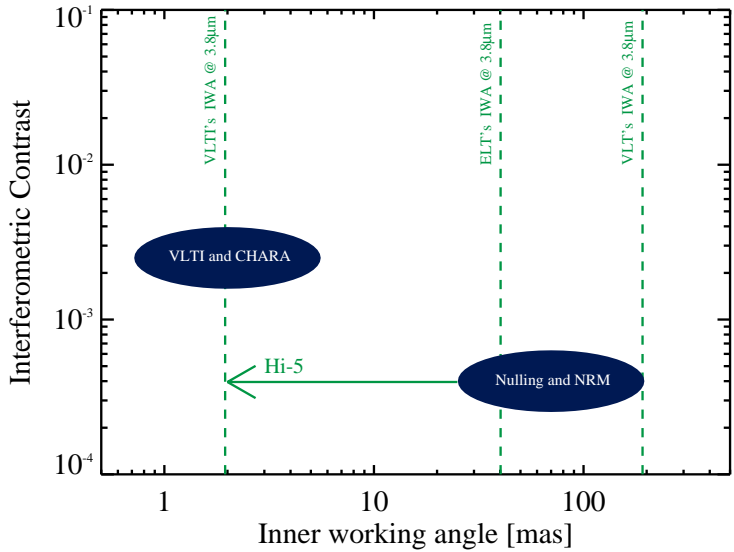}
		\caption{Interferometric contrast as a function of inner working angle for current CHARA and VLTI instruments (i.e., FLUOR, MIRC, PIONIER, and GRAVITY) compared to that of nulling interferometers installed in the Northern hemisphere (i.e., PFN, DRAGONFLY, and LBTI) and non-redundant aperture masking (NRM) experiments (e.g., SPHERE/SAM). From left to right, the vertical dashed lines represent the inner working angle at 3.8\,$\upmu$m of the VLTI, the ELT, and the VLT (computed as $0.5\times\lambda/b$ for the interferometers and as $2\times\lambda/D$ for the single-aperture instruments).}\label{fig1}
		\vspace{-1.5em}
	\end{center}
\end{figure}

Direct imaging is a powerful and historically important observing technique in astrophysics. From Galileo's lens to modern telescopes, scientific progress and discoveries have been guided by the development of imaging instruments with constantly improving angular resolution, sensitivity, and contrast. Current imaging instruments installed on 10-m class ground-based AO-assisted telescopes are strongly limited by contrast within a few resolution elements from the central star, typically $10^{-4}$ at the inner working angles (IWA, a few 0\farcs{1}) to $10^{-5}$ at several $\lambda/D$ from the central star (0\farcs{5}-1\farcs{0}, depending on the wavelength). Interferometric instruments can probe smaller spatial scales but at modest contrast (see Figure~\ref{fig1}). For instance, the VLTI achieves contrasts of a few 10$^{-3}$ in the near-infrared (nIR) with PIONIER (H band) and GRAVITY (K band) down to a few milli-arcseconds (mas). Nulling interferometers installed in the Northern hemisphere and non-redundant aperture masking experiments have demonstrated better contrasts of a few 10$^{-4}$ on baselines shorter than those available at the VLTI.\\ 

Developing high-contrast capabilities has long been recognized as one of the top priorities for future interferometric instruments and for the VLTI in particular \citep[e.g.,][]{Lena:2006}. In the early 2000s, pushed by the need to prepare the way for future space-based infrared interferometric missions, a concept for such an instrument was designed and studied in detail for the VLTI \citep{Absil:2006}. This study established the instrumental constraints on fringe tracking and dispersion control to reach a contrast of 10$^{-4}$, approximately one order of magnitude better than what is achievable with the current and second-generation VLTI instrument suite. While this project did not materialize in an actual instrument, the key scientific questions that it intended to address remain, and high-contrast infrared interferometry is still nowadays the best option to answer them. New scientific questions that would benefit from such an instrument have also appeared in the last 10 years, making the case even stronger. Recent developments in VLT adaptive optics \citep{Dorn:2014}, interferometric data reduction \citep[the so-called Nulling Self Calibration or NSC, see][]{Mennesson:2011}, beam combination architecture \citep{Lacour:2014}, and integrated optics \cite[e.g.,][]{Benisty:2009} offer new possibilities to bring the VLTI to the next level of high-contrast observations at small angular separation. The Hi-5 (High-contrast Interferometry up to 5\,$\upmu$m) project has emerged from these developments. Initial studies are being funded by the H2020 OPTICON Joint Research Network and have officially started with a kickoff meeting held in Li\`ege in October 2017.\\


\section{Heritage and lessons learned from other high-contrast long-baseline interferometers}
\label{sec:1}
\subsection{Visibility interferometry}

Several long-baseline infrared interferometers have been used in the past for science cases requiring high-contrast observations: VLTI/VINCI \citep{Kervella:2000}, IOTA/IONIC \citep{Berger:2003}, CHARA/FLUOR \citep{Foresto:2003}, and VLTI/PIONIER \citep{LeBouquin:2011}. Because these instruments rely on absolute calibration, their contrast is directly related to the stability of the instrumental transfer function and, hence, the accuracy of the measured squared visibilities. The typical statistical uncertainty of the visibilities obtained with these instruments in good conditions is of the order of 1\% and averages down to a few 0.1\% after a complete observing sequence. This corresponds to achievable contrasts of a few 10$^{-3}$, which is sufficient to carry out surveys of hot exozodiacal dust \citep[e.g.,][]{Absil:2013,Ertel:2014} or to search for bright stellar companions \citep[e.g.,][]{Sana:2014,Marion:2014}.\\

A key feature of these instruments include the modal filtering provided by either single-mode fibers (VINCI and FLUOR) or integrated optics (IONIC, PIONIER). Another fundamental requirement to achieve high accuracies is to scan the interferogram faster than the atmospheric turbulence (coherence time), which can limit the observations to relatively bright targets in order to maintain the intended accuracy. In general, the contrast achieved by these concepts is/was limited by polarization errors \citep{LeBouquin:2008, Ertel:2014} and chromaticism of the beam combiner \citep{Defrere:2011}. For VLTI/PIONIER, a special calibration procedure has been developed to mitigate the impact of polarization effects that occur in the VLTI optical train (external to PIONIER). This procedure consists in sampling sufficiently well the dependence of the polarization effect in the sky by observing several CAL-SCI-CAL sequences using different calibrators and science targets at a range of sky positions within one night and correcting for the well defined polarization behavior \citep{Ertel:2014}.

\subsection{Closure phase}

The closure phase is an important interferometric observable that is immune to atmospheric piston \citep{Monnier:2000}. It corresponds to the phase of the triple product (or bispectrum). With current long-baseline interferometers such as VLTI/PIONIER and CHARA/MIRC \citep{Monnier:2004}, contrasts of a few 10$^{-3}$ can be achieved using closure phase \citep{LeBouquin:2012,Marion:2014,Gallenne:2016}. Because of the sparse structure of the interferometric point spread function, the number and orientation of interferometric baselines are important to detect faint components. In addition, the angular separation of the component has to be resolved by at least some baselines, and in that case compensated by accurate interferometric measurements. The best median accuracy in closure phase that has been obtained with long-baseline interferometers is $\sim 0.5^\circ$ for both VLTI/PIONIER and CHARA/MIRC in $\sim 30$\,min integration, while the best accuracy for a given measurement is $0.2^\circ$, which seems to be the current limit. This corresponds to contrast of a few 10$^{-3}$ ($\Delta H \sim 6.5$\,mag), which is currently the best detection limit for companions located within 25\,mas \citep{Gallenne:2015, Roettenbacher:2015}. Fundamentally, closure-phase uncertainties at high flux are limited by fringe tracking errors \citep{Ireland:2013}, which for 100\,Hz bandwidths and 100\,nm uncertainties would be 0.002 degrees in 10 minutes. However, many other instrumental challenges are likely to limit closure-phase uncertainties to about 10 times this limit even in an ideal instrument \citep{Greenbaum:2015}. The loss of coherence caused by spectral smearing can also degrade the constrast for wide field-of-view but this effect can be reduced using high-spectral resolution observations (e.g., with VLTI/GRAVITY). 


\subsection{Nulling interferometry}

A logical way to improve the contrast achieved by an interferometer is to suppress the stellar flux, similar to coronagraphy in single-pupil direct imaging, by employing destructive interference. The basic principle of this technique, first proposed by \cite{Bracewell:1978}, is to combine the beams in phase opposition in order to strongly reduce the on-axis starlight while transmitting the flux of off-axis sources located at angular spacings given by odd multiples of 0.5$\lambda/B$ (where $B$ is the distance between the telescope centers). The high-angular resolution information on the observed object is then encoded in the null depth, which is defined as the ratio of the flux measured in destructive interference and that measured in constructive interference. The advantage of obtaining null depth measurements is that they are more robust against many kinds systematic errors than visibility measurements and hence lead to a better accuracy \citep[e.g.,][]{Colavita:2010}.\\

A number of nulling interferometers have been deployed at US observatories over the last twenty years or so, both across single telescopes and as separate aperture interferometers. These include the BracewelL Infrared Nulling Cryostat \citep[BLINC,][]{Hinz:1998c}, the Keck Interferometer Nuller \citep[KIN,][]{Mennesson:2011}, the Palomar Fiber Nuller \citep[PFN,][]{Mennesson:2011}, the Large Binocular Telescope Interferometer \citep[LBTI,][]{Hinz:2016}, and DRAGONFLY/GLINT on Subaru/SCEXAO \citep{Norris:2014}. Working largely at mid-infrared (mIR) wavelengths (8-14~$\mu$m), where dust in the habitable zones of stars is prominent, and where phase instability is more tractable than at shorter wavelengths, a variety of techniques have been demonstrated on-sky, allowing constraints to be set on exozodiacal emission around a number of nearby stars. Much was learned about instrumental limitations over the course of this work, but mid-IR experiments are inevitably limited by the high background radiation in the mid infrared. On the other hand, two on-sky nullers have operated successfully at near-IR wavelengths, where the background is less of an issue, but where phase fluctuations are more problematic. High null depth accuracies were reached with both the PFN and DRAGONFLY/GLINT at these short wavelengths (approaching 10$^{-4}$ in the best case) due to a combination of factors: the ability to use single mode fibers (PFN) or integrated optics (GLINT), the use of the telescope's extreme adaptive optics system as a cross-aperture fringe tracker, and the introduction of a significantly improved technique for null-depth measurement, i.e., null self-calibration (see Section~\ref{sec:nsc}). In this technique, fine null stabilization is abandoned in favor of using the statistics of the null depth fluctuations to separate the instrumental and astrophysical null depth contributions, in what is essentially an interferometric analog of dark speckle techniques. In fact, null self-calibration significantly relaxes the constraints on the terms contributing to the null error budget, such as the intensity and phase balance, and thus allows for a less constrained nuller design. Even so, high symmetry and stability remain the essential starting points for any high-accuracy nulling interferometer.

\section{Science case of VLTI high-contrast interferometry}
\label{sec:2}

\subsection{Planet formation and young giant planets}

\begin{figure}[!t]
	\begin{center}
		\includegraphics[height=5.1 cm]{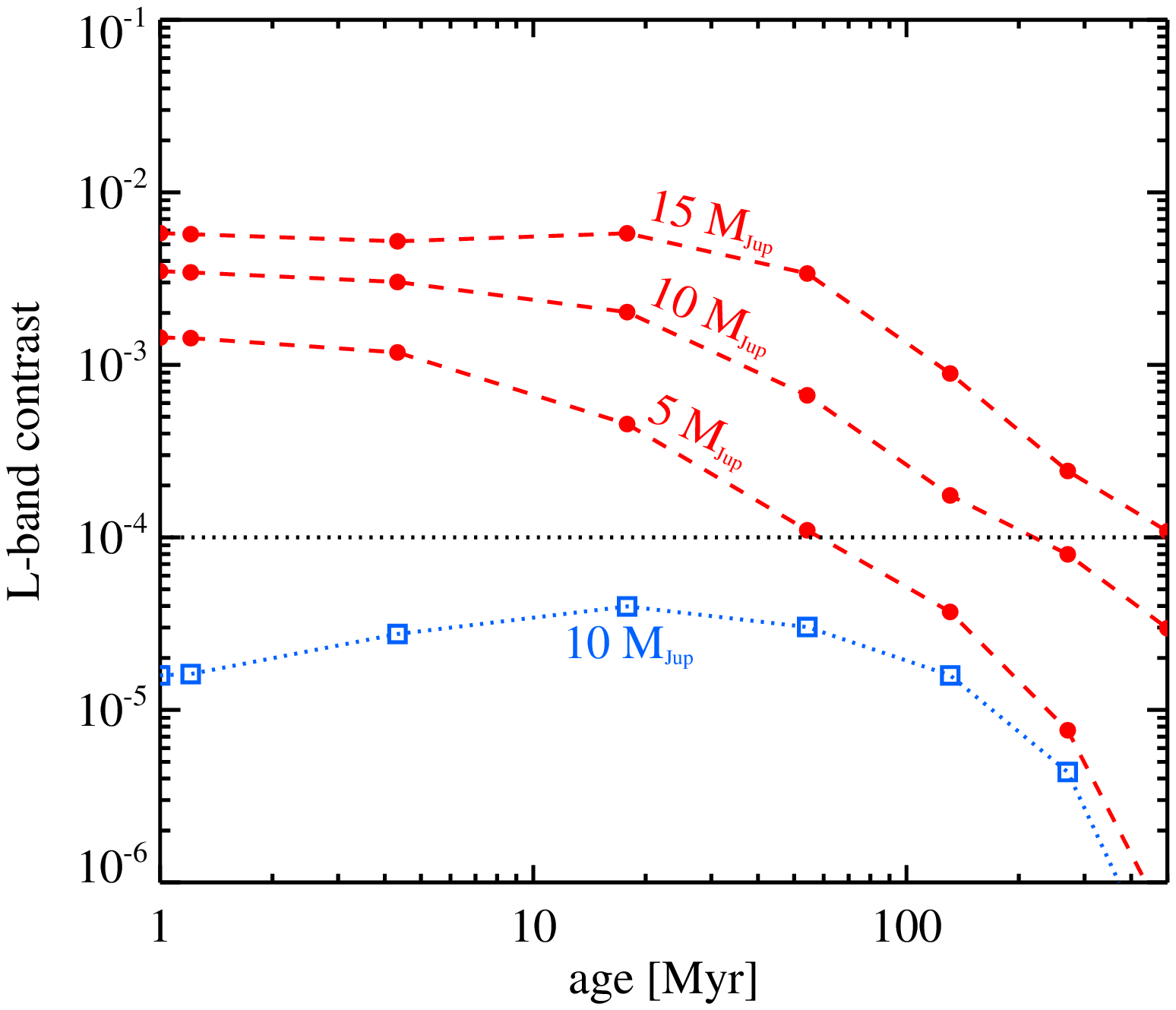}
        \includegraphics[height=5.1 cm]{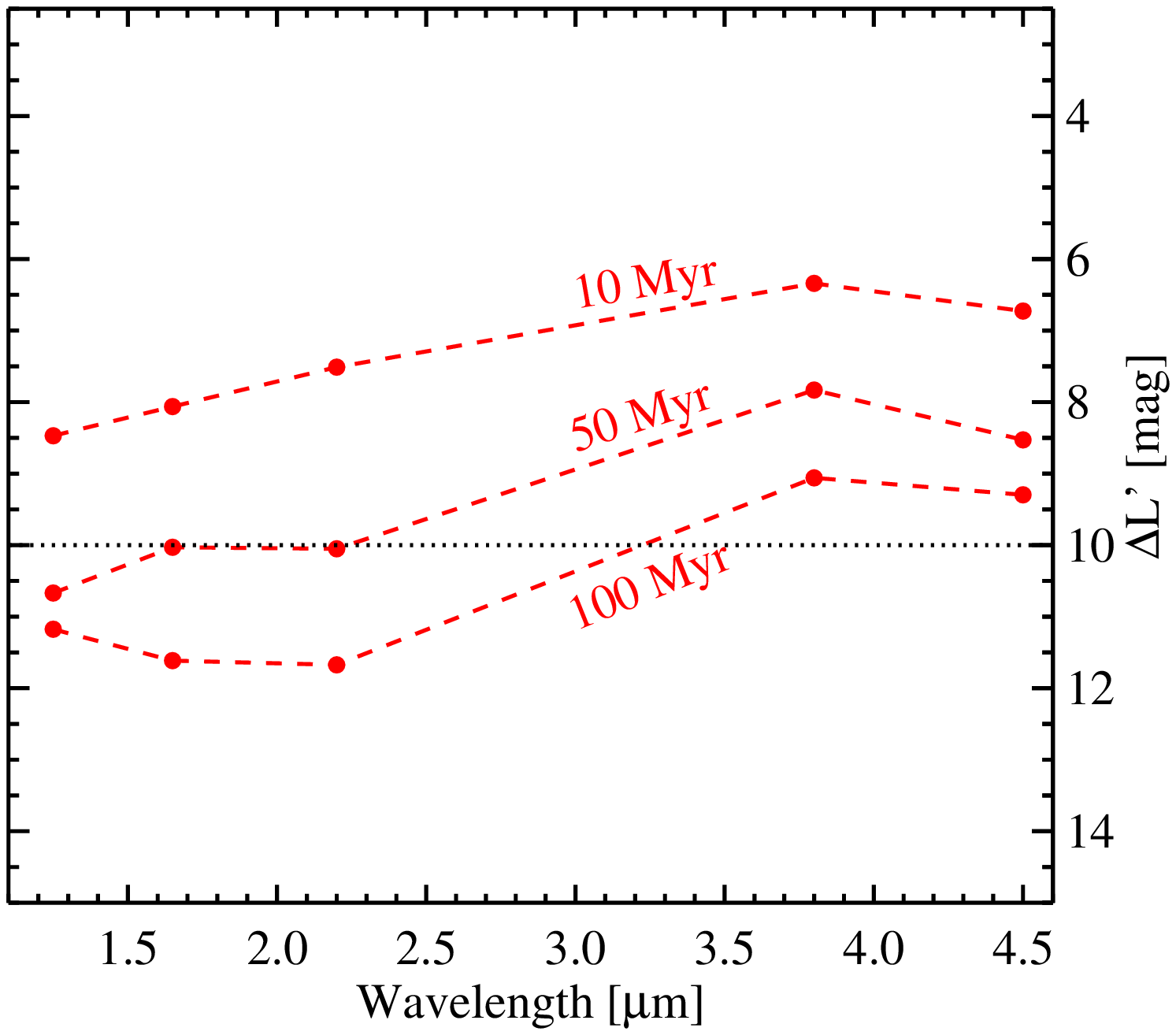}
		\caption{\emph{Left}: Predicted L-band planet-star contrast for a 1-M$_\odot$ star as a function of age and given for different planet masses \citep[red circles, BT-Settl models,][]{Allard:2013}. The lower blue line shows the contrast predicted for the ``cold start" model \citep{Spiegel:2012} and a 10-M$_{\rm Jup}$ planet. \emph{Right}: contrast between a 10-M$_{\rm Jup}$ planet and a 1-M$_\odot$ star as a function of wavelength, showing that the L and M bands provide a better contrast compared to shorter wavelengths and especially for adolescent planets ($\sim$100\,Myr). The dashed horizontal line represents the targeted contrast for Hi-5 (i.e., 10$^{-4}$ or 10 magnitudes). }\label{fig2}
		\vspace{-1.5em}
	\end{center}
\end{figure}

Planets form in the disks around young stars. During the first few million years, these disks are optically thick and the planetary cores are deeply embedded in the disk material. As the planets interact with the disk and the disk dissipates, the planets should become observable through direct imaging. Most planet searches with interferometry in young systems have been conducted using the non-redundant aperture masking (NRM) technique \citep[e.g.][]{Kraus:2012}. However, it has been found that asymmetric emission from the optically thick circumstellar disk can introduce strong phase signals, which can lead to false companion detections \citep[see simulations in][]{Olofsson:2013,Willson:2016}. A high-contrast tIR imager at the VLTI will mitigate these problems by using 10 to 20 times longer baselines than single-aperture NRM interferometry, allowing us to better separate the planet emission from the disk emission. Determining the occurrence rate of giant planets at young age and smaller angular separation can provide critical constraints on planet formation theories and evolution models \citep[e.g.,][]{Spiegel:2012,Mordasini:2012,Allard:2013}. In that regard, the thermal near-infrared (tIR) is a sweet spot to directly detect the photons of self-luminous or irradiated close-in and young ($<$100\,Myr) giant planets (see Figure~\ref{fig2}). Surveys of nearby young stellar moving groups could detect new giant planets at angular distances inaccessible by current instruments and future ELTs. In addition, with a contrast of 10$^{-4}$, previously-known giant exoplanets detected by radial velocity (e.g., $\tau$ Boo b, Gliese 86d) can be resolved and characterized. Low-resolution spectroscopic observations of such planets in the tIR are ideal to derive the radius and effective temperature as well as providing critical information to study the non-equilibrium chemistry of their atmosphere via the CH$_4$ and CO spectral features. The possibility to directly detect rocky planets around nearby low-mass stars (e.g., Proxima b) will also be investigated during the Hi-5 study. 

 
\subsection{Exozodiacal disks}\label{sec:exozodi}

Exozodiacal dust emits primarily in the nIR to mIR where it is outshone by the host star. Due to the small angular scales involved (1\,AU at 10\,pc corresponds to 0.1\,arcsec), the angular resolution required to spatially disentangle the dust from the stellar emission requires the use of interferometry. Thus, exozodis have so far mostly been observed at the CHARA array and the VLTI in the nIR \citep{abs06, Absil:2013, def12b, Ertel:2014, ert16} and at KIN and the LBTI in the mIR \citep{mil11, men13, def15}. These studies give vital statistical insights into the occurrence rates of exozodis as a function of other properties of the systems such as the presence of cold, Kuiper belt-like dust disks or stellar age and spectral type. The main challenge at the moment is linking the nIR and the mIR detections, which critically constrains the systems' architectures and the properties and origin of the dust. However, so far no connection between the detections in the two wavelength ranges has been found. A sensitive instrument operating in the tIR like Hi-5 is the ideal tool to trace the spectral energy distributions of nIR detected exozodis toward longer wavelengths and of mIR detected exozodis toward shorter wavelengths in order to connect the two and to understand non-detections in one wavelength range in the light of detections in the other. Moreover, no sensitive interferometric instrument operating in the thermal infrared is available in the Southern hemisphere so far. MATISSE will not reach the contrast required to detect habitable zone dust at a level comparable to our own zodiacal dust and will thus be limited to the (also very important) characterization of the brightest systems already detected in the nIR. The high contrast mIR capabilities of Hi-5, together with the efficiency increase due to the simultaneous use of four telescopes already demonstrated at the VLTI with PIONIER, will allow for a large survey for habitable zone dust in the Southern hemisphere. This will significantly improve our understanding of the occurrence rates of systems harboring Solar system like exo-zodiacal dust and increase the currently very short list of such systems to be studied in detail.


\subsection{Stellar physics: binarity accross the HR diagram}

Optical interferometry has been used to complement AO assisted imaging survey of stars to estimate the multiplicity fraction. For instance, \cite{Sana:2014} have shown that virtually all massive stars are in multiple system, thanks to a distance-limited survey of close-by massive stars. Extending this result to other class of stars is still to be done, and one of the limitation is contrast: companion detection requires both inner-working angle and detection depth. Apart from multiplicity fraction, another interest of binarity study is determination of fundamental parameters such as dynamical masses or distances. Spectroscopy is currently more sensitive than interferometry, so the stellar mass is only known to the sinus of the inclination of the orbit. Direct imaging and follow up of the companion allows to estimate the true stellar mass. For example, companions around Cepheid pulsating stars are difficult to detect and only the brightest companions are detected using near-infrared interferometry in the 10\,mas separation / 0.005-0.05 contrast regime \citep{Gallenne:2013,Gallenne:2014}. The perspective of spectroscopic radial velocities of the companion (using UV spectroscopy) and interferometric visual orbit opens the possibility for independent distance to Cepheids, rivaling Gaia in term of distance accuracy (Gallenne et al., in preparation). Even if the L band is not optimum to detect hot Cepheids companion, an ten-fold improvement in contrast compared to current H-band instrument will still lead more detections.

\subsection{Extragalactic astrophysics} 

Most of the physical processes in Active Galactic Nuclei (AGN) take place on scales of a few parsec or less (i.e.\ $\lesssim 100 \, \mathrm{mas}$ for the nearest galaxies). Hence, it requires interferometic methods to resolve the relevant scales. 
AGN are ``faint'' for infrared interferometry, with fluxes of $F_\mathrm{K} < 70 \, \mathrm{mJy}$ ($K > 10 \, \mathrm{mag}$) in the near-infrared, and $F_\mathrm{N} < \, 1\mathrm{Jy}$ ($N > 4 \, \mathrm{mag}$, with a few exceptions) in the mid-infrared. Additionally, AGN spectra are very red and they often appear extended in the optical, leading to limitations for fringe tracking and poor AO correction. Nevertheless, interferometry of several AGN with the VLTI and the Keck Interferometer has shown that their dust distributions are compact, with sizes roughly scaling with the square root of the intrinsic luminosity \citep[e.g.][]{Tristram:2011, Kishimoto:2011, Burtscher:2013}. The few better resolved sources reveal a two component structure, with a central disk and an emission extending in the polar direction \citep[e.g.][]{Hoenig:2012, Tristram:2014, Lopez-Gonzaga:2016}. However, the number of AGN observable by current instruments is very limited and most sources only appear marginally resolved, especially towards shorter wavelengths. Further progress in our understanding of AGN can hence only be expected from an instrument providing \emph{high accuracy visibility measurements} as well as a \emph{high sensitivity}. This will allow to better constrain larger samples of marginally resolved AGN, especially if also the ATs on the longest baselines can be used. Furthermore, by combining interferometric with reverberation measurements, direct distances to such sources can be determined \citep{Hoenig:2014}, with the possibility to independently constrain the Hubble parameter.

\section{Improving the contrast of the VLTI}\label{sec:3}

\subsection{Integrated optics}

Integrated optics (IO) is a key component of current high-contrast VLTI interferometers such as PIONIER (H band) and GRAVITY (K band). In the tIR, recent efforts have been targeting the development of components with ultrafast laser inscription in mid-infrared-transparent glasses. Ultrafast laser inscription (ULI) is a versatile technique using highly focused pulses from a femtosecond laser to induce permanent structural modifications in a large variety of glasses \citep[see][]{Gattass:2008}. The modifications are responsible for localised changes of the refractive index, which can be used to manufacture photonic devices based on waveguides. Remarkably, three dimensional structures can be written by scanning the glass samples under the laser focus. Particularly interesting for tIR interferometry is the processing of chalcogenide glasses such as Gallium Lanthanum Sulfide \citep{Rodenas:2012} or Germanium Arsenic Sulfide \citep{DAmico:2014}, which have transparency windows extending to a wavelength of about 10~$\upmu$m. Waveguides with a core-cladding contrast of several $10^{-3}$ and propagation losses at the 0.7-0.9 dB/cm level are routinely manufactured with ULI techniques. Photonic building blocks such as Y-junctions \citep{Rodenas:2012} and 2x2 directional couplers have been demonstrated \citep{Arriola:2014}. Couplers similar to the latter component were recently characterised in the L-, L$^{\prime}$- and M-bands, demonstrating high broadband contrasts, low spectral phase distortion, and 30\% to 60\% measured throughput \citep{Tepper:2017a,Tepper:2017b}. More advanced components, allowing the combination of several telescopes, have also been manufactured and tested. The first component was a 3-telescope N-band combiner based on cascaded Y-junctions with three-dimensional avoidance of waveguide cross-overs \citep{Rodenas:2012}. More recently, a 2-telescope ABCD combination unit \citep{Benisty:2009} and a 4-telescope beam combiner based on Discrete Beam Combiner geometry \citep{Minardi:2010} were manufactured with ULI and tested interferometrically with monochromatic light at 3.39~$\upmu$m \citep{Diener:2017}. Both components showed that retrieval of complex visibilities with high signal to noise is possible at relatively low illumination levels (about 1000 counts per combined channel and 10 counts of readout noise rms).\\

An alternative to ultrafast laser inscription to fabricate integrated optics beam combiners is the use of classical methods, such as Ti:indiffusion inside electro-optic crystals. These waveguides are interesting as the refractive index of the material, and therefore the phase of the propagating optical beam, can be modified by the application of an external voltage. In the particular case of Lithium Niobate crystals, the transparency window reaches 5.2~$\upmu$m allowing to cover L and M bands. Using this technology, phase and intensity modulators \citep{Heidmann:2012}, achieving on-chip fringe scanning, fringe locking and high contrast interferometry (36dB) have been demonstrated at 3.39~$\upmu$m \citep{Martin:2014a}. Concepts such as active 2T ABCD \citep{Heidmann:2011} and 3T AC \citep{Martin:2014b} infrared beam combiners have been validated experimentally. However, propagation losses in these systems are currently too high (typ. 5 dB/cm). Therefore, novel methods such as ULI presented above, but developed in electro-optic crystals, are being tested for waveguide fabrication, showing low propagation losses (1.5 dB/cm) in the first prototypes \citep{Nguyen:2017}. Finally, note that two-dimensional photolithography on a platform with Ge,As,Se and Ge,As,S based glasses offsets the potential for less than 0.5dB/cm losses in mm-scale chips tolerant of low bend-radius \citep{Kenchington:2017}.

\subsection{Fringe tracking}

Phase-referenced interferometers require accurate and robust fringe tracking for sensitive background-limited observations and high-contrast imaging. Reaching contrasts of a few 10$^{-4}$ at 3.8~$\upmu$m puts however strong constraints on the fringe tracker which has to deliver closed-loop optical path difference (OPD) residuals of a few nanometers RMS \citep{Serabyn:2000, Absil:2006}. For nulling, these constraints can be relaxed by using advanced data reduction techniques (see Section~\ref{sec:nsc}) but it is currently not clear whether this technique can be used to produce high-contrast images. A promising technique currently under investigation uses a dual fringe tracking and low-order adaptive optics concept based on a combination of non-redundant aperture interferometry and eigen-phase in asymmetric pupil wavefront sensing \citep[e.g.,][]{Martinache:2016}. Applied to the VLTI with the NAOMI adaptive optics system functioning on the ATs, fringe tracking sensitivity of H=12.6 in most seeing conditions can in theory be achieved, which is not matched by conventional techniques.\\
 
Another limiting factor of current high-contrast long-baseline interferometers is the phase chromaticism induced by random fluctuations in the water vapor differential column density above each aperture (or water vapor seeing). This component of the OPD is not correctly tracked at the wavelength of the science channel when this one operates at a wavelength different from that of the fringe sensor. The impact of this effect on infrared interferometry has been addressed extensively in the literature, either in a general context \citep{Colavita:2004} or applied to specific instruments that include phase-referenced modes using K-band light such as VLTI/MIDI \citep{Meisner:2003,Matter:2010,Pott:2012}, the KIN \citep{Colavita:2010}, and the LBTI \citep{Defrere:2016}. This effect will be seriously considered in the context of the Hi-5 study. 

\subsection{Limiting magnitude}

One important metric of an interferometer is its limiting magnitude, which is related to different instrumental parameters such as aperture size, throughput, and scan speed. For instance, the limiting magnitude of VLTI/PIONIER is constrained towards faint targets because the fringes are tracked internally on the science data and the scan speed needs to be large enough compared to the coherence time in order to minimize the degrading effects of atmospheric turbulence during a scan. In order to track the fringes, at such short integration times the (correlated) flux must be high enough to reach a good signal-to-noise ratio in each single scan. The problem is similar for phase-referenced instruments, which suffer from significant performance degradation for faint stars because a slower acquisition rate has to be used. One way to relax the constraints on scan speed is to use real-time data from accelerometers attached to the optical train \citep{Bohm:2017} and/or adaptive optics wave-front sensing \citep{Pott:2016}. Improving the limiting magnitude is crucial for various reasons and, in particular, for increasing the sample of observable young stars and extragalactic objects.

\subsection{Data reduction technique}\label{sec:nsc}

Data reduction is another very important aspect to consider when optimizing the design of an interferometer. For nulling interferometers, the use of classical data reduction approaches imposes strict constraints on the co-phasing accuracy and intensity mismatch achieved by the instrument in order to reach contrasts of a few $10^{-4}$ \citep[e.g.,][]{Absil:2006}. A new statistical method has recently shown that it is possible to reach contrasts of a few $10^{-4}$ with significantly relaxed instrumental constraints \citep[][]{Hanot:2011,Mennesson:2011}. For instance, a contrast of a few $10^{-4}$ has been achieved with the LBTI despite closed-loop OPD residuals of approximately 400\,nm RMS \citep{Defrere:2016}. This method, currently only applicable to two-telescope interferometers, has to be generalized for more telescopes in order to be used with the VLTI. The effects of lower effective duty cycle on null depth and the effect of low signal-to-noise within the fringe tracker inverse bandwidth also has to be investigated.  

\subsection{Other possibilities}

In additions to the elements described above, other promising avenues to improve the contrast of the VLTI need to be investigated. For instance, a new interferometer architecture combining nulling with phase closure measurements has recently been proposed \citep{Lacour:2014}. This design consists of a nulling stage and a set of ABCD beam combiners which combine the nulled outputs of the preceding stage, with the goal of characterizing the coherence of the remaining light in a manner robust against imperfect cophasing of the incoming stellar light. Another promising way to improve the contrast of the VLTI is to combine interferometry with high-dispersion spectroscopy such as performed with single-aperture telescopes \citep[e.g.,][]{Snellen:2015}. This technique is currently being explored with VLTI/GRAVITY and can be further extended to be used with a nulling instrument. Finally, one can also consider an interesting new idea of symmetric beam combination scheme that is insensitive to polarization states \citep[i.e., the Cross Cuber Nuller,][]{Henault:2014}. The application of these techniques to the VLTI will be investigated during the Hi-5 study.

\section{Synergies with other instruments}\label{sec:synergies}

Hi-5 will be complementary to several future high-angular resolution instruments operating in the tIR as described below.

\begin{itemize}
\item MATISSE \citep[Multi AperTure mid-Infrared SpectroScopic Experiment,][]{Lopez:2014, Matter:2016a} is the second-generation tIR and mIR spectrograph and imager for the VLTI. MATISSE will provide a wide wavelength coverage, from 2.8 to 13~$\upmu$m, associated with a milli-arcsecond scale angular resolution (3 mas in L band; 10 mas in N band), and various spectral resolutions from R $\sim$30 to R$\sim$5000. In terms of performance, theoretical MATISSE visibility accuracies of 1 to 3 percent in L and M bands, and 8 percent in N band, were estimated for a 20~Jy source; estimates based on SNR calculations including the contribution of the fundamental noises (source photon noise, readout noise, thermal background photon noise) and the transfer function variations \citep[][]{Matter:2016b}. More recently, in the frame of the MATISSE test phase in lab, many instrumental visibilities were measured over 4 hours in LM band, and 3 days in N band. Those measurements were performed with a very bright artificial IR source in order to estimate the instrumental contribution to the accuracy without being limited by the fundamental noises. Such a source would have an equivalent flux, if observed with the UTs, of 20 to 70 Jy in N-band, 400 Jy in M-band, and 600 Jy in L band. This lead to absolute visibility accuracies lower than 0.5 percent in L band, 0.4 percent in M band, and 2.5 percent in N band, on average over the corresponding spectral band. Those promising results are extensively described in internal ESO documents written by the MATISSE consortium (private communication with A.~Matter). Eventually, the on-sky tests (commissioning), starting in March 2018, will provide the real on-sky performance (sensitivity, accuracy) of MATISSE, which will notably include the effects of the sky thermal background fluctuations, the atmospheric turbulence, and the on-sky calibration. 

\item ELT/METIS \citep[][]{Brandl:2016} is the Mid-infrared E-ELT Imager and Spectro-graph for the European Extremely Large Telescope. METIS will provide diffraction limited imaging and medium resolution slit spectroscopy in the 3 to 19~$\upmu$m range, as well as high resolution (R = 100000) integral field spectroscopy from 2.9 to 5.3~$\upmu$m. Assuming a collecting aperture of 39\,m in diameter, METIS will provide an angular resolution in the nIR similar to that of Hi-5 in the tIR (see Figure~\ref{fig1}). VLTI/Hi-5 will hence provide complementary high-contrast observations to characterize the observed planets and circumstellar disks. In particular, a VLTI instrument can make use of less-solicited telescopes such as the ATs to follow-up in the tIR new ELT/METIS discoveries. 

\item PFI \citep[Planet Formation Imager,][]{Monnier:2016,Kraus:2016,Ireland:2016} is currently a science-driven, international initiative to develop the roadmap for a future ground-based facility that will be optimised to image planet-forming disks on the spatial scale where the protoplanets are assembled, which is the Hill sphere of the forming planets. The goal of PFI will be to detect and characterise protoplanets during their first $\sim 100$ million years and trace how the planet population changes due to migration processes, unveiling the processes that determine the final architecture of exoplanetary systems. With $\sim 20$ telescope elements and baselines of $\sim 3$~km, the PFI concept is optimised for imaging complex scenes at tIR and mIR wavelengths (3-12$\upmu$m) and at 0.1 milliarcsecond resolution. Hence, Hi-5's mission will be ``explorative", while PFI's mission will be to provide a comprehensive picture of planet formation and characterisation (resolving circumplanetary disks). Hi-5 and PFI will also share many common technology challenges, for instance on tIR beam combination, accurate/robust fringe tracking, and nulling schemes.

\item FKSI \citep[Fourier-Kelvin Stellar Interferometer,][]{Danchi:2003} is a concept for a small structurally connected space-based infrared interferometer, with a 12.5-m baseline, operating from 3 to 8\,$\upmu$m or possible 10\,$\upmu$m, passively cooled to 60\,K, operating primarily in a nulling (starlight suppressing) mode for the detection and characterization of exoplanets, debris disks, and extrasolar zodiacal dust levels. It would have the highest angular resolution of any infrared space instrument ever made with a nominal resolution of 40\,mas at a 5\,$\upmu$m center wavelength. This resolution exceeds that of Spitzer by a factor of 38 and JWST by a factor of 5. Relatively little work has been done since 2010 on FKSI, due to funding limitations. However, within the past year or so there has been renewed interest at NASA regarding missions with a lifecycle cost of less than one billion dollars. In addition, there is increasing interest at NASA for distributed spacecraft mission concepts, as well as novel low-cost mission concepts, either for a specific astrophysics observation/measurement or to advance technologies, with some science. Hi-5 and FKSI will share common technology challenges such as tIR beam combination and accurate/robust fringe tracking. 

\end{itemize}

\section{Summary and conclusions}

The VLTI currently achieves contrasts of a few 10$^{-3}$ in the near-infrared and second-generation instruments are not designed to do better. Achieving deeper contrasts at small inner working angles is however mandatory to make scientific progress in various fields of astrophysics and, in particular, in exoplanet science. On the VLTI, gaining one order of magnitude (i.e., contrasts of at least a few 10$^{-4}$) is today within reach as demonstrated with ground-based nulling interferometers in the northern hemisphere and non-redundant aperture masking instruments. In addition, a key technology that made the success of PIONIER (H band) and GRAVITY (K band) is now coming to maturity for the the thermal near-infrared (L and M bands), a sweet spot to image young giant exoplanets. New ideas have also emerged to improve the contrast of long-baseline interferometers (i.e., combining nulling and closure phase, advanced fringe tracking, high-dispersion interferometry). These new possibilities for high-contrast imaging will be investigated in the context of the Hi-5 study, which will particularly explore the limits of exoplanet detection from the ground with existing interferometers. Besides the clear scientific motivation, a new high-contrast VLTI imaging instrument will serve as a key technology demonstrator for future major interferometric instruments such as PFI and TPF-I/DARWIN-like missions. Technology demonstration will include fringe tracking, advanced beam combination strategies, thermal near-infrared integrated optics components (which greatly reduce the complexity of the instrument), and four-telescope statistical data reduction. 

\begin{acknowledgements}
The authors acknowledge the support from the H2020 OPTICON Joint Research Network. DD and OA thank the Belgian national funds for scientific research (FNRS). SK acknowledges support from an ERC Starting Grant (Grant Agreement No.\ 639889) and STFC Rutherford Fellowship (ST/J004030/1).
\end{acknowledgements}

\bibliographystyle{spbasic}      

\begin{thebibliography}{84}
\providecommand{\natexlab}[1]{#1}
\providecommand{\url}[1]{{#1}}
\providecommand{\urlprefix}{URL }
\expandafter\ifx\csname urlstyle\endcsname\relax
  \providecommand{\doi}[1]{DOI~\discretionary{}{}{}#1}\else
  \providecommand{\doi}{DOI~\discretionary{}{}{}\begingroup
  \urlstyle{rm}\Url}\fi
\providecommand{\eprint}[2][]{\url{#2}}

\bibitem[{{Absil} et~al(2006{\natexlab{a}}){Absil}, {den Hartog}, {Gondoin},
  {Fabry}, {Wilhelm}, {Gitton}, and {Puech}}]{Absil:2006}
{Absil} O, {den Hartog} R, {Gondoin} P, {Fabry} P, {Wilhelm} R, {Gitton} P,
  {Puech} F (2006{\natexlab{a}}) {Performance study of ground-based infrared
  Bracewell interferometers. Application to the detection of exozodiacal dust
  disks with GENIE}. \aap 448:787--800, \doi{10.1051/0004-6361:20053516},
  \eprint{astro-ph/0511223}

\bibitem[{{Absil} et~al(2006{\natexlab{b}}){Absil}, {di Folco}, {M{\'e}rand},
  {Augereau}, {Coud{\'e} du Foresto}, {Aufdenberg}, {Kervella}, {Ridgway},
  {Berger}, {ten Brummelaar}, {Sturmann}, {Sturmann}, {Turner}, and
  {McAlister}}]{abs06}
{Absil} O, {di Folco} E, {M{\'e}rand} A, {Augereau} JC, {Coud{\'e} du Foresto}
  V, {Aufdenberg} JP, {Kervella} P, {Ridgway} ST, {Berger} DH, {ten Brummelaar}
  TA, {Sturmann} J, {Sturmann} L, {Turner} NH, {McAlister} HA
  (2006{\natexlab{b}}) {Circumstellar material in the <ASTROBJ>Vega</ASTROBJ>
  inner system revealed by CHARA/FLUOR}. \aap 452:237--244,
  \doi{10.1051/0004-6361:20054522}, \eprint{arXiv:astro-ph/0604260}

\bibitem[{{Absil} et~al(2013){Absil}, {Defr{\`e}re}, {Coud{\'e} du Foresto},
  {Di Folco}, {M{\'e}rand}, {Augereau}, {Ertel}, {Hanot}, {Kervella},
  {Mollier}, {Scott}, {Che}, {Monnier}, {Thureau}, {Tuthill}, {ten Brummelaar},
  {McAlister}, {Sturmann}, {Sturmann}, and {Turner}}]{Absil:2013}
{Absil} O, {Defr{\`e}re} D, {Coud{\'e} du Foresto} V, {Di Folco} E,
  {M{\'e}rand} A, {Augereau} JC, {Ertel} S, {Hanot} C, {Kervella} P, {Mollier}
  B, {Scott} N, {Che} X, {Monnier} JD, {Thureau} N, {Tuthill} PG, {ten
  Brummelaar} TA, {McAlister} HA, {Sturmann} J, {Sturmann} L, {Turner} N (2013)
  {A near-infrared interferometric survey of debris-disc stars. III. First
  statistics based on 42 stars observed with CHARA/FLUOR}. \aap 555:A104,
  \doi{10.1051/0004-6361/201321673}, \eprint{1307.2488}

\bibitem[{{Allard} et~al(2013){Allard}, {Homeier}, {Freytag}, {Schaffenberger},
  {}, and {Rajpurohit}}]{Allard:2013}
{Allard} F, {Homeier} D, {Freytag} B, {Schaffenberger}, {} W, {Rajpurohit} AS
  (2013) {Progress in modeling very low mass stars, brown dwarfs, and planetary
  mass objects.} Memorie della Societa Astronomica Italiana Supplementi 24:128,
  \eprint{1302.6559}

\bibitem[{{Arriola} et~al(2014){Arriola}, {Mukherjee}, {Choudhury}, {Labadie},
  and {Thomson}}]{Arriola:2014}
{Arriola} A, {Mukherjee} S, {Choudhury} D, {Labadie} L, {Thomson} RR (2014)
  {Ultrafast laser inscription of mid-IR directional couplers for stellar
  interferometry}. Optics Letters 39:4820, \doi{10.1364/OL.39.004820},
  \eprint{1408.5953}

\bibitem[{{Benisty} et~al(2009){Benisty}, {Berger}, {Jocou}, {Labeye},
  {Malbet}, {Perraut}, and {Kern}}]{Benisty:2009}
{Benisty} M, {Berger} JP, {Jocou} L, {Labeye} P, {Malbet} F, {Perraut} K,
  {Kern} P (2009) {An integrated optics beam combiner for the second generation
  VLTI instruments}. \aap 498:601--613, \doi{10.1051/0004-6361/200811083},
  \eprint{0902.2442}

\bibitem[{{Berger} et~al(2003){Berger}, {Haguenauer}, {Kern},
  {Rousselet-Perraut}, {Malbet}, {Gluck}, {Lagny}, {Schanen-Duport}, {Laurent},
  {Delboulbe}, {Tatulli}, {Traub}, {Carleton}, {Millan-Gabet}, {Monnier},
  {Pedretti}, and {Ragland}}]{Berger:2003}
{Berger} JP, {Haguenauer} P, {Kern} PY, {Rousselet-Perraut} K, {Malbet} F,
  {Gluck} S, {Lagny} L, {Schanen-Duport} I, {Laurent} E, {Delboulbe} A,
  {Tatulli} E, {Traub} WA, {Carleton} N, {Millan-Gabet} R, {Monnier} JD,
  {Pedretti} E, {Ragland} S (2003) {An integrated-optics 3-way beam combiner
  for IOTA}. In: {Traub} WA (ed) Interferometry for Optical Astronomy II,
  \procspie, vol 4838, pp 1099--1106, \doi{10.1117/12.457983}

\bibitem[{B\"{o}hm et~al(2017)B\"{o}hm, Pott, Kürster, Sawodny, Defrère, and
  Hinz}]{Bohm:2017}
B\"{o}hm M, Pott JU, Kürster M, Sawodny O, Defrère D, Hinz P (2017) Delay
  compensation for real time disturbance estimation at extremely large
  telescopes. IEEE Transactions on Control Systems Technology 25(4):1384--1393,
  \doi{10.1109/TCST.2016.2601627}

\bibitem[{{Bracewell}(1978)}]{Bracewell:1978}
{Bracewell} RN (1978) {Detecting nonsolar planets by spinning infrared
  interferometer}. \nat 274:780--+

\bibitem[{{Brandl} et~al(2016){Brandl}, {Ag{\'o}cs}, {Aitink-Kroes}, {Bertram},
  {Bettonvil}, {van Boekel}, {Boulade}, {Feldt}, {Glasse}, {Glauser},
  {G{\"u}del}, {Hurtado}, {Jager}, {Kenworthy}, {Mach}, {Meisner}, {Meyer},
  {Pantin}, {Quanz}, {Schmid}, {Stuik}, {Veninga}, and
  {Waelkens}}]{Brandl:2016}
{Brandl} BR, {Ag{\'o}cs} T, {Aitink-Kroes} G, {Bertram} T, {Bettonvil} F, {van
  Boekel} R, {Boulade} O, {Feldt} M, {Glasse} A, {Glauser} A, {G{\"u}del} M,
  {Hurtado} N, {Jager} R, {Kenworthy} MA, {Mach} M, {Meisner} J, {Meyer} M,
  {Pantin} E, {Quanz} S, {Schmid} HM, {Stuik} R, {Veninga} A, {Waelkens} C
  (2016) {Status of the mid-infrared E-ELT imager and spectrograph METIS}. In:
  Ground-based and Airborne Instrumentation for Astronomy VI, \procspie, vol
  9908, p 990820, \doi{10.1117/12.2233974}

\bibitem[{{Burtscher} et~al(2013){Burtscher}, {Meisenheimer}, {Tristram},
  {Jaffe}, {H{\"o}nig}, {Davies}, {Kishimoto}, {Pott}, {R{\"o}ttgering},
  {Schartmann}, {Weigelt}, and {Wolf}}]{Burtscher:2013}
{Burtscher} L, {Meisenheimer} K, {Tristram} KRW, {Jaffe} W, {H{\"o}nig} SF,
  {Davies} RI, {Kishimoto} M, {Pott} JU, {R{\"o}ttgering} H, {Schartmann} M,
  {Weigelt} G, {Wolf} S (2013) {A diversity of dusty AGN tori. Data release for
  the VLTI/MIDI AGN Large Program and first results for 23 galaxies}. \aap
  558:A149, \doi{10.1051/0004-6361/201321890}, \eprint{1307.2068}

\bibitem[{{Colavita} et~al(2004){Colavita}, {Swain}, {Akeson}, {Koresko}, and
  {Hill}}]{Colavita:2004}
{Colavita} MM, {Swain} MR, {Akeson} RL, {Koresko} CD, {Hill} RJ (2004) {Effects
  of Atmospheric Water Vapor on Infrared Interferometry}. \pasp 116:876--885,
  \doi{10.1086/424472}

\bibitem[{{Colavita} et~al(2010){Colavita}, {Serabyn}, {Ragland},
  {Millan-Gabet}, and {Akeson}}]{Colavita:2010}
{Colavita} MM, {Serabyn} E, {Ragland} S, {Millan-Gabet} R, {Akeson} RL (2010)
  {Keck Interferometer nuller instrument performance}. In: Society of
  Photo-Optical Instrumentation Engineers (SPIE) Conference Series, Society of
  Photo-Optical Instrumentation Engineers (SPIE) Conference Series, vol 7734,
  p~0, \doi{10.1117/12.857166}

\bibitem[{{Coud{\'e} du Foresto} et~al(2003){Coud{\'e} du Foresto}, {Borde},
  {Merand}, {Baudouin}, {Remond}, {Perrin}, {Ridgway}, {ten Brummelaar}, and
  {McAlister}}]{Foresto:2003}
{Coud{\'e} du Foresto} V, {Borde} PJ, {Merand} A, {Baudouin} C, {Remond} A,
  {Perrin} GS, {Ridgway} ST, {ten Brummelaar} TA, {McAlister} HA (2003) {FLUOR
  fibered beam combiner at the CHARA array}. In: {Traub} WA (ed) Interferometry
  for Optical Astronomy II, \procspie, vol 4838, pp 280--285,
  \doi{10.1117/12.459942}

\bibitem[{{D'Amico} et~al(2014){D'Amico}, {Cheng}, {Mauclair}, {Troles},
  {Calvez}, {Nazabal}, {Caillaud}, {Martin}, {Arezki}, {LeCoarer}, {Kern}, and
  {Stoian}}]{DAmico:2014}
{D'Amico} C, {Cheng} G, {Mauclair} C, {Troles} J, {Calvez} L, {Nazabal} V,
  {Caillaud} C, {Martin} G, {Arezki} B, {LeCoarer} E, {Kern} P, {Stoian} R
  (2014) {Large-mode-area infrared guiding in ultrafast laser written
  waveguides in Sulfur-based chalcogenide glasses}. Optics Express 22:13,091,
  \doi{10.1364/OE.22.013091}

\bibitem[{{Danchi} et~al(2003){Danchi}, {Deming}, {Kuchner}, and
  {Seager}}]{Danchi:2003}
{Danchi} WC, {Deming} D, {Kuchner} MJ, {Seager} S (2003) {Detection of Close-In
  Extrasolar Giant Planets Using theFourier-Kelvin Stellar Interferometer}.
  \apjl 597:L57--L60, \doi{10.1086/379640}, \eprint{astro-ph/0309361}

\bibitem[{{Defr{\`e}re} et~al(2011){Defr{\`e}re}, {Absil}, {Augereau}, {di
  Folco}, {Berger}, {Coud{\'e} du Foresto}, {Kervella}, {Le Bouquin},
  {Lebreton}, {Millan-Gabet}, {Monnier}, {Olofsson}, and
  {Traub}}]{Defrere:2011}
{Defr{\`e}re} D, {Absil} O, {Augereau} JC, {di Folco} E, {Berger} JP,
  {Coud{\'e} du Foresto} V, {Kervella} P, {Le Bouquin} JB, {Lebreton} J,
  {Millan-Gabet} R, {Monnier} JD, {Olofsson} J, {Traub} W (2011) {Hot
  exozodiacal dust resolved around Vega with IOTA/IONIC}. \aap 534:A5,
  \doi{10.1051/0004-6361/201117017}, \eprint{1108.3698}

\bibitem[{{Defr{\`e}re} et~al(2012){Defr{\`e}re}, {Lebreton}, {Le Bouquin},
  {Lagrange}, {Absil}, {Augereau}, {Berger}, {di Folco}, {Ertel}, {Kluska},
  {Montagnier}, {Millan-Gabet}, {Traub}, and {Zins}}]{def12b}
{Defr{\`e}re} D, {Lebreton} J, {Le Bouquin} JB, {Lagrange} AM, {Absil} O,
  {Augereau} JC, {Berger} JP, {di Folco} E, {Ertel} S, {Kluska} J, {Montagnier}
  G, {Millan-Gabet} R, {Traub} W, {Zins} G (2012) {Hot circumstellar material
  resolved around <ASTROBJ>{$\beta$} Pic</ASTROBJ> with VLTI/PIONIER}. \aap
  546:L9, \doi{10.1051/0004-6361/201220287}, \eprint{1210.1914}

\bibitem[{{Defr{\`e}re} et~al(2015){Defr{\`e}re}, {Hinz}, {Skemer}, {Kennedy},
  {Bailey}, {Hoffmann}, {Mennesson}, {Millan-Gabet}, {Danchi}, {Absil}, {Arbo},
  {Beichman}, {Brusa}, {Bryden}, {Downey}, {Durney}, {Esposito}, {Gaspar},
  {Grenz}, {Haniff}, {Hill}, {Lebreton}, {Leisenring}, {Males}, {Marion},
  {McMahon}, {Montoya}, {Morzinski}, {Pinna}, {Puglisi}, {Rieke}, {Roberge},
  {Serabyn}, {Sosa}, {Stapeldfeldt}, {Su}, {Vaitheeswaran}, {Vaz},
  {Weinberger}, and {Wyatt}}]{def15}
{Defr{\`e}re} D, {Hinz} PM, {Skemer} AJ, {Kennedy} GM, {Bailey} VP, {Hoffmann}
  WF, {Mennesson} B, {Millan-Gabet} R, {Danchi} WC, {Absil} O, {Arbo} P,
  {Beichman} C, {Brusa} G, {Bryden} G, {Downey} EC, {Durney} O, {Esposito} S,
  {Gaspar} A, {Grenz} P, {Haniff} C, {Hill} JM, {Lebreton} J, {Leisenring} JM,
  {Males} JR, {Marion} L, {McMahon} TJ, {Montoya} M, {Morzinski} KM, {Pinna} E,
  {Puglisi} A, {Rieke} G, {Roberge} A, {Serabyn} E, {Sosa} R, {Stapeldfeldt} K,
  {Su} K, {Vaitheeswaran} V, {Vaz} A, {Weinberger} AJ, {Wyatt} MC (2015)
  {First-light LBT Nulling Interferometric Observations: Warm Exozodiacal Dust
  Resolved within a Few AU of {$\eta$} Crv}. \apj 799:42,
  \doi{10.1088/0004-637X/799/1/42}, \eprint{1501.04144}

\bibitem[{{Defr{\`e}re} et~al(2016){Defr{\`e}re}, {Hinz}, {Mennesson},
  {Hoffmann}, {Millan-Gabet}, {Skemer}, {Bailey}, {Danchi}, {Downey}, {Durney},
  {Grenz}, {Hill}, {McMahon}, {Montoya}, {Spalding}, {Vaz}, {Absil}, {Arbo},
  {Bailey}, {Brusa}, {Bryden}, {Esposito}, {Gaspar}, {Haniff}, {Kennedy},
  {Leisenring}, {Marion}, {Nowak}, {Pinna}, {Powell}, {Puglisi}, {Rieke},
  {Roberge}, {Serabyn}, {Sosa}, {Stapeldfeldt}, {Su}, {Weinberger}, and
  {Wyatt}}]{Defrere:2016}
{Defr{\`e}re} D, {Hinz} PM, {Mennesson} B, {Hoffmann} WF, {Millan-Gabet} R,
  {Skemer} AJ, {Bailey} V, {Danchi} WC, {Downey} EC, {Durney} O, {Grenz} P,
  {Hill} JM, {McMahon} TJ, {Montoya} M, {Spalding} E, {Vaz} A, {Absil} O,
  {Arbo} P, {Bailey} H, {Brusa} G, {Bryden} G, {Esposito} S, {Gaspar} A,
  {Haniff} CA, {Kennedy} GM, {Leisenring} JM, {Marion} L, {Nowak} M, {Pinna} E,
  {Powell} K, {Puglisi} A, {Rieke} G, {Roberge} A, {Serabyn} E, {Sosa} R,
  {Stapeldfeldt} K, {Su} K, {Weinberger} AJ, {Wyatt} MC (2016) {Nulling Data
  Reduction and On-sky Performance of the Large Binocular Telescope
  Interferometer}. \apj 824:66, \doi{10.3847/0004-637X/824/2/66},
  \eprint{1601.06866}

\bibitem[{Diener et~al(2017)Diener, Tepper, Labadie, Pertsch, Nolte, and
  Minardi}]{Diener:2017}
Diener R, Tepper J, Labadie L, Pertsch T, Nolte S, Minardi S (2017) Towards
  3d-photonic, multi-telescope beam combiners for mid-infrared
  astrointerferometry. Opt Express 25(16):19,262--19,274,
  \doi{10.1364/OE.25.019262},
  \urlprefix\url{http://www.opticsexpress.org/abstract.cfm?URI=oe-25-16-19262}

\bibitem[{{Dorn} et~al(2014){Dorn}, {Aller-Carpentier}, {Andolfato}, {Berger},
  {Delplancke-Str{\"o}bele}, {Dupuy}, {Fedrigo}, {Gitton}, {Hubin}, {Le
  Louarn}, {Lilley}, {Jolley}, {Marchetti}, {Mclay}, {Paufique}, {Pasquini},
  {Quentin}, {Rakich}, {Ridings}, {Reyes}, {Schmid}, {Suarez}, {Phan}, and
  {Woillez}}]{Dorn:2014}
{Dorn} RJ, {Aller-Carpentier} E, {Andolfato} L, {Berger} JP,
  {Delplancke-Str{\"o}bele} F, {Dupuy} C, {Fedrigo} E, {Gitton} P, {Hubin} N,
  {Le Louarn} M, {Lilley} P, {Jolley} P, {Marchetti} E, {Mclay} S, {Paufique}
  J, {Pasquini} L, {Quentin} J, {Rakich} A, {Ridings} R, {Reyes} J, {Schmid} C,
  {Suarez} M, {Phan} DT, {Woillez} J (2014) {NAOMI {\mdash} A New Adaptive
  Optics Module for Interferometry}. The Messenger 156:12--15

\bibitem[{{Ertel} et~al(2014){Ertel}, {Absil}, {Defr{\`e}re}, {Le Bouquin},
  {Augereau}, {Marion}, {Blind}, {Bonsor}, {Bryden}, {Lebreton}, and
  {Milli}}]{Ertel:2014}
{Ertel} S, {Absil} O, {Defr{\`e}re} D, {Le Bouquin} JB, {Augereau} JC, {Marion}
  L, {Blind} N, {Bonsor} A, {Bryden} G, {Lebreton} J, {Milli} J (2014) {A
  near-infrared interferometric survey of debris-disk stars. IV. An unbiased
  sample of 92 southern stars observed in H band with VLTI/PIONIER}. \aap
  570:A128, \doi{10.1051/0004-6361/201424438}, \eprint{1409.6143}

\bibitem[{{Ertel} et~al(2016){Ertel}, {Defr{\`e}re}, {Absil}, {Le Bouquin},
  {Augereau}, {Berger}, {Blind}, {Bonsor}, {Lagrange}, {Lebreton}, {Marion},
  {Milli}, and {Olofsson}}]{ert16}
{Ertel} S, {Defr{\`e}re} D, {Absil} O, {Le Bouquin} JB, {Augereau} JC, {Berger}
  JP, {Blind} N, {Bonsor} A, {Lagrange} AM, {Lebreton} J, {Marion} L, {Milli}
  J, {Olofsson} J (2016) {A near-infrared interferometric survey of debris-disc
  stars. V. PIONIER search for variability}. \aap 595:A44,
  \doi{10.1051/0004-6361/201527721}, \eprint{1608.05731}

\bibitem[{{Gallenne} et~al(2013){Gallenne}, {Monnier}, {M{\'e}rand},
  {Kervella}, {Kraus}, {Schaefer}, {Gieren}, {Pietrzy{\'n}ski}, {Szabados},
  {Che}, {Baron}, {Pedretti}, {McAlister}, {ten Brummelaar}, {Sturmann},
  {Sturmann}, {Turner}, {Farrington}, and {Vargas}}]{Gallenne:2013}
{Gallenne} A, {Monnier} JD, {M{\'e}rand} A, {Kervella} P, {Kraus} S, {Schaefer}
  GH, {Gieren} W, {Pietrzy{\'n}ski} G, {Szabados} L, {Che} X, {Baron} F,
  {Pedretti} E, {McAlister} H, {ten Brummelaar} T, {Sturmann} J, {Sturmann} L,
  {Turner} N, {Farrington} C, {Vargas} N (2013) {Multiplicity of Galactic
  Cepheids from long-baseline interferometry. I. CHARA/MIRC detection of the
  companion of V1334 Cygni}. \aap 552:A21, \doi{10.1051/0004-6361/201321091},
  \eprint{1302.1817}

\bibitem[{{Gallenne} et~al(2014){Gallenne}, {M{\'e}rand}, {Kervella},
  {Breitfelder}, {Le Bouquin}, {Monnier}, {Gieren}, {Pilecki}, and
  {Pietrzy{\'n}ski}}]{Gallenne:2014}
{Gallenne} A, {M{\'e}rand} A, {Kervella} P, {Breitfelder} J, {Le Bouquin} JB,
  {Monnier} JD, {Gieren} W, {Pilecki} B, {Pietrzy{\'n}ski} G (2014)
  {Multiplicity of Galactic Cepheids from long-baseline interferometry. II. The
  Companion of AX Circini revealed with VLTI/PIONIER}. \aap 561:L3,
  \doi{10.1051/0004-6361/201322883}, \eprint{1312.1950}

\bibitem[{{Gallenne} et~al(2015){Gallenne}, {M{\'e}rand}, {Kervella},
  {Monnier}, {Schaefer}, {Baron}, {Breitfelder}, {Le Bouquin}, {Roettenbacher},
  {Gieren}, {Pietrzy{\'n}ski}, {McAlister}, {ten Brummelaar}, {Sturmann},
  {Sturmann}, {Turner}, {Ridgway}, and {Kraus}}]{Gallenne:2015}
{Gallenne} A, {M{\'e}rand} A, {Kervella} P, {Monnier} JD, {Schaefer} GH,
  {Baron} F, {Breitfelder} J, {Le Bouquin} JB, {Roettenbacher} RM, {Gieren} W,
  {Pietrzy{\'n}ski} G, {McAlister} H, {ten Brummelaar} T, {Sturmann} J,
  {Sturmann} L, {Turner} N, {Ridgway} S, {Kraus} S (2015) {Robust high-contrast
  companion detection from interferometric observations. The CANDID algorithm
  and an application to six binary Cepheids}. \aap 579:A68,
  \doi{10.1051/0004-6361/201525917}, \eprint{1505.02715}

\bibitem[{{Gallenne} et~al(2016){Gallenne}, {M{\'e}rand}, {Kervella},
  {Monnier}, {Schaefer}, {Roettenbacher}, {Gieren}, {Pietrzy{\'n}ski},
  {McAlister}, {ten Brummelaar}, {Sturmann}, {Sturmann}, {Turner}, and
  {Anderson}}]{Gallenne:2016}
{Gallenne} A, {M{\'e}rand} A, {Kervella} P, {Monnier} JD, {Schaefer} GH,
  {Roettenbacher} RM, {Gieren} W, {Pietrzy{\'n}ski} G, {McAlister} H, {ten
  Brummelaar} T, {Sturmann} J, {Sturmann} L, {Turner} N, {Anderson} RI (2016)
  {Multiplicity of Galactic Cepheids from long-baseline interferometry - III.
  Sub-percent limits on the relative brightness of a close companion of
  {$\delta$} Cephei}. \mnras 461:1451--1456, \doi{10.1093/mnras/stw1375},
  \eprint{1606.01108}

\bibitem[{{Gattass} and {Mazur}(2008)}]{Gattass:2008}
{Gattass} RR, {Mazur} E (2008) {Femtosecond laser micromachining in transparent
  materials}. Nature Photonics 2:219--225, \doi{10.1038/nphoton.2008.47}

\bibitem[{{Greenbaum} et~al(2015){Greenbaum}, {Pueyo}, {Sivaramakrishnan}, and
  {Lacour}}]{Greenbaum:2015}
{Greenbaum} AZ, {Pueyo} L, {Sivaramakrishnan} A, {Lacour} S (2015) {An
  Image-plane Algorithm for JWST's Non-redundant Aperture Mask Data}. \apj
  798:68, \doi{10.1088/0004-637X/798/2/68}, \eprint{1411.3446}

\bibitem[{{Hanot} et~al(2011){Hanot}, {Mennesson}, {Martin}, {Liewer}, {Loya},
  {Mawet}, {Riaud}, {Absil}, and {Serabyn}}]{Hanot:2011}
{Hanot} C, {Mennesson} B, {Martin} S, {Liewer} K, {Loya} F, {Mawet} D, {Riaud}
  P, {Absil} O, {Serabyn} E (2011) {Improving Interferometric Null Depth
  Measurements using Statistical Distributions: Theory and First Results with
  the Palomar Fiber Nuller}. \apj 729:110, \doi{10.1088/0004-637X/729/2/110},
  \eprint{1103.4719}

\bibitem[{{Heidmann} et~al(2011){Heidmann}, {Caballero}, {Nolot}, {Gineys},
  {Moulin}, {Delboulb{\'e}}, {Jocou}, {Le Bouquin}, {Berger}, and
  {Martin}}]{Heidmann:2011}
{Heidmann} S, {Caballero} O, {Nolot} A, {Gineys} M, {Moulin} T, {Delboulb{\'e}}
  A, {Jocou} L, {Le Bouquin} JB, {Berger} JP, {Martin} G (2011) {Two telescopes
  ABCD electro-optic beam combiner based on lithium niobate for near infrared
  stellar interferometry}. In: Nonlinear Optics and Applications V, \procspie,
  vol 8071, p 807108, \doi{10.1117/12.886725}

\bibitem[{{Heidmann} et~al(2012){Heidmann}, {Courjal}, and
  {Martin}}]{Heidmann:2012}
{Heidmann} S, {Courjal} N, {Martin} G (2012) {Double polarization active Y
  junctions in the L band, based on Ti:LiNbO\_3 lithium niobate waveguides:
  polarization and contrast performances}. Optics Letters 37:3318,
  \doi{10.1364/OL.37.003318}

\bibitem[{{H{\'e}nault} and {Spang}(2014)}]{Henault:2014}
{H{\'e}nault} F, {Spang} A (2014) {Cheapest nuller in the world: crossed
  beamsplitter cubes}. In: Optical and Infrared Interferometry IV, \procspie,
  vol 9146, p 914604, \doi{10.1117/12.2055091}, \eprint{1407.2719}

\bibitem[{{Hinz} et~al(1998){Hinz}, {Angel}, {Hoffmann}, {McCarthy}, {McGuire},
  {Cheselka}, {Hora}, and {Woolf}}]{Hinz:1998c}
{Hinz} PM, {Angel} JRP, {Hoffmann} WF, {McCarthy} DW, {McGuire} PC, {Cheselka}
  M, {Hora} JL, {Woolf} NJ (1998) {Imaging circumstellar environments with a
  nulling interferometer}. \nat 395:251--253, \doi{10.1038/26172}

\bibitem[{{Hinz} et~al(2016){Hinz}, {Defr{\`e}re}, {Skemer}, {Bailey}, {Stone},
  {Spalding}, {Vaz}, {Pinna}, {Puglisi}, {Esposito}, {Montoya}, {Downey},
  {Leisenring}, {Durney}, {Hoffmann}, {Hill}, {Millan-Gabet}, {Mennesson},
  {Danchi}, {Morzinski}, {Grenz}, {Skrutskie}, and {Ertel}}]{Hinz:2016}
{Hinz} PM, {Defr{\`e}re} D, {Skemer} A, {Bailey} V, {Stone} J, {Spalding} E,
  {Vaz} A, {Pinna} E, {Puglisi} A, {Esposito} S, {Montoya} M, {Downey} E,
  {Leisenring} J, {Durney} O, {Hoffmann} W, {Hill} J, {Millan-Gabet} R,
  {Mennesson} B, {Danchi} W, {Morzinski} K, {Grenz} P, {Skrutskie} M, {Ertel} S
  (2016) {Overview of LBTI: a multipurpose facility for high spatial resolution
  observations}. In: Optical and Infrared Interferometry and Imaging V,
  \procspie, vol 9907, p 990704, \doi{10.1117/12.2233795}

\bibitem[{{H{\"o}nig} et~al(2012){H{\"o}nig}, {Kishimoto}, {Antonucci},
  {Marconi}, {Prieto}, {Tristram}, and {Weigelt}}]{Hoenig:2012}
{H{\"o}nig} SF, {Kishimoto} M, {Antonucci} R, {Marconi} A, {Prieto} MA,
  {Tristram} K, {Weigelt} G (2012) {Parsec-scale Dust Emission from the Polar
  Region in the Type 2 Nucleus of NGC 424}. \apj 755:149,
  \doi{10.1088/0004-637X/755/2/149}, \eprint{1206.4307}

\bibitem[{{H{\"o}nig} et~al(2014){H{\"o}nig}, {Watson}, {Kishimoto}, and
  {Hjorth}}]{Hoenig:2014}
{H{\"o}nig} SF, {Watson} D, {Kishimoto} M, {Hjorth} J (2014) {A dust-parallax
  distance of 19 megaparsecs to the supermassive black hole in NGC 4151}. \nat
  515:528--530, \doi{10.1038/nature13914}, \eprint{1411.7032}

\bibitem[{{Ireland}(2013)}]{Ireland:2013}
{Ireland} MJ (2013) {Phase errors in diffraction-limited imaging: contrast
  limits for sparse aperture masking}. \mnras 433:1718--1728,
  \doi{10.1093/mnras/stt859}, \eprint{1301.6205}

\bibitem[{{Ireland} et~al(2016){Ireland}, {Monnier}, {Kraus}, {Isella},
  {Minardi}, {Petrov}, {ten Brummelaar}, {Young}, {Vasisht}, {Mozurkewich},
  {Rinehart}, {Michael}, {van Belle}, and {Woillez}}]{Ireland:2016}
{Ireland} MJ, {Monnier} JD, {Kraus} S, {Isella} A, {Minardi} S, {Petrov} R,
  {ten Brummelaar} T, {Young} J, {Vasisht} G, {Mozurkewich} D, {Rinehart} S,
  {Michael} EA, {van Belle} G, {Woillez} J (2016) {Status of the Planet
  Formation Imager (PFI) concept}. In: Optical and Infrared Interferometry and
  Imaging V, \procspie, vol 9907, p 99071L, \doi{10.1117/12.2233926},
  \eprint{1608.00582}

\bibitem[{{Kenchington Goldsmith} et~al(2017){Kenchington Goldsmith},
  {Cvetojevic}, {Ireland}, and {Madden}}]{Kenchington:2017}
{Kenchington Goldsmith} HD, {Cvetojevic} N, {Ireland} M, {Madden} S (2017)
  {Fabrication tolerant chalcogenide mid-infrared multimode interference
  coupler design with applications for Bracewell nulling interferometry}.
  Optics Express 25:3038, \doi{10.1364/OE.25.003038}, \eprint{1702.00468}

\bibitem[{{Kervella} et~al(2000){Kervella}, {Coud{\'e} du Foresto},
  {Glindemann}, and {Hofmann}}]{Kervella:2000}
{Kervella} P, {Coud{\'e} du Foresto} V, {Glindemann} A, {Hofmann} R (2000)
  {VINCI: the VLT Interferometer commissioning instrument}. In: {L{\'e}na} P,
  {Quirrenbach} A (eds) Interferometry in Optical Astronomy, \procspie, vol
  4006, pp 31--42, \doi{10.1117/12.390227}

\bibitem[{{Kishimoto} et~al(2011){Kishimoto}, {H{\"o}nig}, {Antonucci},
  {Barvainis}, {Kotani}, {Tristram}, {Weigelt}, and {Levin}}]{Kishimoto:2011}
{Kishimoto} M, {H{\"o}nig} SF, {Antonucci} R, {Barvainis} R, {Kotani} T,
  {Tristram} KRW, {Weigelt} G, {Levin} K (2011) {The innermost dusty structure
  in active galactic nuclei as probed by the Keck interferometer}. \aap
  527:A121, \doi{10.1051/0004-6361/201016054}, \eprint{1012.5359}

\bibitem[{{Kraus} and {Ireland}(2012)}]{Kraus:2012}
{Kraus} AL, {Ireland} MJ (2012) {LkCa 15: A Young Exoplanet Caught at
  Formation?} \apj 745:5, \doi{10.1088/0004-637X/745/1/5}, \eprint{1110.3808}

\bibitem[{{Kraus} et~al(2016){Kraus}, {Monnier}, {Ireland}, {Duch{\^e}ne},
  {Espaillat}, {H{\"o}nig}, {Juhasz}, {Mordasini}, {Olofsson}, {Paladini},
  {Stassun}, {Turner}, {Vasisht}, {Harries}, {Bate}, {Gonzalez}, {Matter},
  {Zhu}, {Panic}, {Regaly}, {Morbidelli}, {Meru}, {Wolf}, {Ilee}, {Berger},
  {Zhao}, {Kral}, {Morlok}, {Bonsor}, {Ciardi}, {Kane}, {Kratter}, {Laughlin},
  {Pepper}, {Raymond}, {Labadie}, {Nelson}, {Weigelt}, {ten Brummelaar},
  {Pierens}, {Oudmaijer}, {Kley}, {Pope}, {Jensen}, {Bayo}, {Smith},
  {Boyajian}, {Quiroga-Nu{\~n}ez}, {Millan-Gabet}, {Chiavassa}, {Gallenne},
  {Reynolds}, {de Wit}, {Wittkowski}, {Millour}, {Gandhi}, {Ramos Almeida},
  {Alonso Herrero}, {Packham}, {Kishimoto}, {Tristram}, {Pott}, {Surdej},
  {Buscher}, {Haniff}, {Lacour}, {Petrov}, {Ridgway}, {Tuthill}, {van Belle},
  {Armitage}, {Baruteau}, {Benisty}, {Bitsch}, {Paardekooper}, {Pinte},
  {Masset}, and {Rosotti}}]{Kraus:2016}
{Kraus} S, {Monnier} JD, {Ireland} MJ, {Duch{\^e}ne} G, {Espaillat} C,
  {H{\"o}nig} S, {Juhasz} A, {Mordasini} C, {Olofsson} J, {Paladini} C,
  {Stassun} K, {Turner} N, {Vasisht} G, {Harries} TJ, {Bate} MR, {Gonzalez} JF,
  {Matter} A, {Zhu} Z, {Panic} O, {Regaly} Z, {Morbidelli} A, {Meru} F, {Wolf}
  S, {Ilee} J, {Berger} JP, {Zhao} M, {Kral} Q, {Morlok} A, {Bonsor} A,
  {Ciardi} D, {Kane} SR, {Kratter} K, {Laughlin} G, {Pepper} J, {Raymond} S,
  {Labadie} L, {Nelson} RP, {Weigelt} G, {ten Brummelaar} T, {Pierens} A,
  {Oudmaijer} R, {Kley} W, {Pope} B, {Jensen} ELN, {Bayo} A, {Smith} M,
  {Boyajian} T, {Quiroga-Nu{\~n}ez} LH, {Millan-Gabet} R, {Chiavassa} A,
  {Gallenne} A, {Reynolds} M, {de Wit} WJ, {Wittkowski} M, {Millour} F,
  {Gandhi} P, {Ramos Almeida} C, {Alonso Herrero} A, {Packham} C, {Kishimoto}
  M, {Tristram} KRW, {Pott} JU, {Surdej} J, {Buscher} D, {Haniff} C, {Lacour}
  S, {Petrov} R, {Ridgway} S, {Tuthill} P, {van Belle} G, {Armitage} P,
  {Baruteau} C, {Benisty} M, {Bitsch} B, {Paardekooper} SJ, {Pinte} C, {Masset}
  F, {Rosotti} G (2016) {Planet Formation Imager (PFI): science vision and key
  requirements}. In: Optical and Infrared Interferometry and Imaging V,
  \procspie, vol 9907, p 99071K, \doi{10.1117/12.2231067}, \eprint{1608.00578}

\bibitem[{{Lacour} et~al(2014){Lacour}, {Tuthill}, {Monnier}, {Kotani},
  {Gauchet}, and {Labeye}}]{Lacour:2014}
{Lacour} S, {Tuthill} P, {Monnier} JD, {Kotani} T, {Gauchet} L, {Labeye} P
  (2014) {A new interferometer architecture combining nulling with phase
  closure measurements}. \mnras 439:4018--4029, \doi{10.1093/mnras/stu258},
  \eprint{1306.5184}

\bibitem[{{Le Bouquin} and {Absil}(2012)}]{LeBouquin:2012}
{Le Bouquin} JB, {Absil} O (2012) {On the sensitivity of closure phases to
  faint companions in optical long baseline interferometry}. \aap 541:A89,
  \doi{10.1051/0004-6361/201117891}, \eprint{1204.3721}

\bibitem[{{Le Bouquin} et~al(2008){Le Bouquin}, {Rousselet-Perraut}, {Berger},
  {Herwats}, {Benisty}, {Absil}, {Defrere}, {Monnier}, and
  {Traub}}]{LeBouquin:2008}
{Le Bouquin} JB, {Rousselet-Perraut} K, {Berger} JP, {Herwats} E, {Benisty} M,
  {Absil} O, {Defrere} D, {Monnier} J, {Traub} W (2008) {Polar-interferometry:
  what can be learnt from the IOTA/IONIC experiment}. In: Optical and Infrared
  Interferometry, \procspie, vol 7013, p 70130F, \doi{10.1117/12.786377}

\bibitem[{{Le Bouquin} et~al(2011){Le Bouquin}, {Berger}, {Lazareff}, {Zins},
  {Haguenauer}, {Jocou}, {Kern}, {Millan-Gabet}, {Traub}, {Absil}, {Augereau},
  {Benisty}, {Blind}, {Bonfils}, {Bourget}, {Delboulbe}, {Feautrier},
  {Germain}, {Gitton}, {Gillier}, {Kiekebusch}, {Kluska}, {Knudstrup},
  {Labeye}, {Lizon}, {Monin}, {Magnard}, {Malbet}, {Maurel}, {M{\'e}nard},
  {Micallef}, {Michaud}, {Montagnier}, {Morel}, {Moulin}, {Perraut}, {Popovic},
  {Rabou}, {Rochat}, {Rojas}, {Roussel}, {Roux}, {Stadler}, {Stefl}, {Tatulli},
  and {Ventura}}]{LeBouquin:2011}
{Le Bouquin} JB, {Berger} JP, {Lazareff} B, {Zins} G, {Haguenauer} P, {Jocou}
  L, {Kern} P, {Millan-Gabet} R, {Traub} W, {Absil} O, {Augereau} JC, {Benisty}
  M, {Blind} N, {Bonfils} X, {Bourget} P, {Delboulbe} A, {Feautrier} P,
  {Germain} M, {Gitton} P, {Gillier} D, {Kiekebusch} M, {Kluska} J, {Knudstrup}
  J, {Labeye} P, {Lizon} JL, {Monin} JL, {Magnard} Y, {Malbet} F, {Maurel} D,
  {M{\'e}nard} F, {Micallef} M, {Michaud} L, {Montagnier} G, {Morel} S,
  {Moulin} T, {Perraut} K, {Popovic} D, {Rabou} P, {Rochat} S, {Rojas} C,
  {Roussel} F, {Roux} A, {Stadler} E, {Stefl} S, {Tatulli} E, {Ventura} N
  (2011) {PIONIER: a 4-telescope visitor instrument at VLTI}. \aap 535:A67,
  \doi{10.1051/0004-6361/201117586}, \eprint{1109.1918}

\bibitem[{{L\'ena} et~al(2005){L\'ena}, {Absil}, {Borkowski}, {Herwats}, D.,
  S., and P.}]{Lena:2006}
{L\'ena} P, {Absil} O, {Borkowski} V, {Herwats} E, D M, S Q, P R (2005) {37th
  Liège International Astrophysics Colloquium : conclusions and perspectives}.
  Bulletin de la Société Royale des Sciences de Li\`ege 74:203--229

\bibitem[{{Lopez} et~al(2014){Lopez}, {Lagarde}, {Jaffe}, {Petrov},
  {Sch{\"o}ller}, {Antonelli}, {Beckmann}, {Berio}, {Bettonvil}, {Glindemann},
  {Gonzalez}, {Graser}, {Hofmann}, {Millour}, {Robbe-Dubois}, {Venema}, {Wolf},
  {Henning}, {Lanz}, {Weigelt}, {Agocs}, {Bailet}, {Bresson}, {Bristow},
  {Dugu{\'e}}, {Heininger}, {Kroes}, {Laun}, {Lehmitz}, {Neumann}, {Augereau},
  {Avila}, {Behrend}, {van Belle}, {Berger}, {van Boekel}, {Bonhomme},
  {Bourget}, {Brast}, {Clausse}, {Connot}, {Conzelmann}, {Cruzal{\`e}bes},
  {Csepany}, {Danchi}, {Delbo}, {Delplancke}, {Dominik}, {van Duin}, {Elswijk},
  {Fantei}, {Finger}, {Gabasch}, {Gay}, {Girard}, {Girault}, {Gitton},
  {Glazenborg}, {Gont{\'e}}, {Guitton}, {Guniat}, {De Haan}, {Haguenauer},
  {Hanenburg}, {Hogerheijde}, {ter Horst}, {Hron}, {Hugues}, {Hummel},
  {Idserda}, {Ives}, {Jakob}, {Jasko}, {Jolley}, {Kiraly}, {K{\"o}hler},
  {Kragt}, {Kroener}, {Kuindersma}, {Labadie}, {Leinert}, {Le Poole}, {Lizon},
  {Lucuix}, {Marcotto}, {Martinache}, {Martinot-Lagarde}, {Mathar}, {Matter},
  {Mauclert}, {Mehrgan}, {Meilland}, {Meisenheimer}, {Meisner}, {Mellein},
  {Menardi}, {Menut}, {Merand}, {Morel}, {Mosoni}, {Navarro}, {Nussbaum},
  {Ottogalli}, {Palsa}, {Panduro}, {Pantin}, {Parra}, {Percheron}, {Duc},
  {Pott}, {Pozna}, {Przygodda}, {Rabbia}, {Richichi}, {Rigal}, {Roelfsema},
  {Rupprecht}, {Schertl}, {Schmidt}, {Schuhler}, {Schuil}, {Spang},
  {Stegmeier}, {Thiam}, {Tromp}, {Vakili}, {Vannier}, {Wagner}, and
  {Woillez}}]{Lopez:2014}
{Lopez} B, {Lagarde} S, {Jaffe} W, {Petrov} R, {Sch{\"o}ller} M, {Antonelli} P,
  {Beckmann} U, {Berio} P, {Bettonvil} F, {Glindemann} A, {Gonzalez} JC,
  {Graser} U, {Hofmann} KH, {Millour} F, {Robbe-Dubois} S, {Venema} L, {Wolf}
  S, {Henning} T, {Lanz} T, {Weigelt} G, {Agocs} T, {Bailet} C, {Bresson} Y,
  {Bristow} P, {Dugu{\'e}} M, {Heininger} M, {Kroes} G, {Laun} W, {Lehmitz} M,
  {Neumann} U, {Augereau} JC, {Avila} G, {Behrend} J, {van Belle} G, {Berger}
  JP, {van Boekel} R, {Bonhomme} S, {Bourget} P, {Brast} R, {Clausse} JM,
  {Connot} C, {Conzelmann} R, {Cruzal{\`e}bes} P, {Csepany} G, {Danchi} W,
  {Delbo} M, {Delplancke} F, {Dominik} C, {van Duin} A, {Elswijk} E, {Fantei}
  Y, {Finger} G, {Gabasch} A, {Gay} J, {Girard} P, {Girault} V, {Gitton} P,
  {Glazenborg} A, {Gont{\'e}} F, {Guitton} F, {Guniat} S, {De Haan} M,
  {Haguenauer} P, {Hanenburg} H, {Hogerheijde} M, {ter Horst} R, {Hron} J,
  {Hugues} Y, {Hummel} C, {Idserda} J, {Ives} D, {Jakob} G, {Jasko} A, {Jolley}
  P, {Kiraly} S, {K{\"o}hler} R, {Kragt} J, {Kroener} T, {Kuindersma} S,
  {Labadie} L, {Leinert} C, {Le Poole} R, {Lizon} JL, {Lucuix} C, {Marcotto} A,
  {Martinache} F, {Martinot-Lagarde} G, {Mathar} R, {Matter} A, {Mauclert} N,
  {Mehrgan} L, {Meilland} A, {Meisenheimer} K, {Meisner} J, {Mellein} M,
  {Menardi} S, {Menut} JL, {Merand} A, {Morel} S, {Mosoni} L, {Navarro} R,
  {Nussbaum} E, {Ottogalli} S, {Palsa} R, {Panduro} J, {Pantin} E, {Parra} T,
  {Percheron} I, {Duc} TP, {Pott} JU, {Pozna} E, {Przygodda} F, {Rabbia} Y,
  {Richichi} A, {Rigal} F, {Roelfsema} R, {Rupprecht} G, {Schertl} D, {Schmidt}
  C, {Schuhler} N, {Schuil} M, {Spang} A, {Stegmeier} J, {Thiam} L, {Tromp} N,
  {Vakili} F, {Vannier} M, {Wagner} K, {Woillez} J (2014) {An Overview of the
  MATISSE Instrument {\mdash} Science, Concept and Current Status}. The
  Messenger 157:5--12

\bibitem[{{L{\'o}pez-Gonzaga} et~al(2016){L{\'o}pez-Gonzaga}, {Burtscher},
  {Tristram}, {Meisenheimer}, and {Schartmann}}]{Lopez-Gonzaga:2016}
{L{\'o}pez-Gonzaga} N, {Burtscher} L, {Tristram} KRW, {Meisenheimer} K,
  {Schartmann} M (2016) {Mid-infrared interferometry of 23 AGN tori: On the
  significance of polar-elongated emission}. \aap 591:A47,
  \doi{10.1051/0004-6361/201527590}, \eprint{1602.05592}

\bibitem[{{Marion} et~al(2014){Marion}, {Absil}, {Ertel}, {Le Bouquin},
  {Augereau}, {Blind}, {Defr{\`e}re}, {Lebreton}, and {Milli}}]{Marion:2014}
{Marion} L, {Absil} O, {Ertel} S, {Le Bouquin} JB, {Augereau} JC, {Blind} N,
  {Defr{\`e}re} D, {Lebreton} J, {Milli} J (2014) {Searching for faint
  companions with VLTI/PIONIER. II. 92 main sequence stars from the Exozodi
  survey}. \aap 570:A127, \doi{10.1051/0004-6361/201424780}, \eprint{1409.6105}

\bibitem[{{Martin} et~al(2014{\natexlab{a}}){Martin}, {Heidmann}, {Rauch},
  {Jocou}, and {Courjal}}]{Martin:2014a}
{Martin} G, {Heidmann} S, {Rauch} JY, {Jocou} L, {Courjal} N
  (2014{\natexlab{a}}) {Electro-optic fringe locking and photometric tuning
  using a two-stage Mach-Zehnder lithium niobate waveguide for high-contrast
  mid-infrared interferometry}. Optical Engineering 53(3):034101,
  \doi{10.1117/1.OE.53.3.034101}

\bibitem[{{Martin} et~al(2014{\natexlab{b}}){Martin}, {Heidmann}, {Thomas}, {de
  Mengin}, {Jocou}, {Ulliac}, {Courjal}, {Morand}, {Benech}, and {le
  Coarer}}]{Martin:2014b}
{Martin} G, {Heidmann} S, {Thomas} F, {de Mengin} M, {Jocou} L, {Ulliac} G,
  {Courjal} N, {Morand} A, {Benech} P, {le Coarer} EP (2014{\natexlab{b}})
  {Lithium Niobate active beam combiners: results of on-chip fringe locking,
  fringe scanning and high contrast integrated optics interferometry and
  spectrometry}. In: Optical and Infrared Interferometry IV, \procspie, vol
  9146, p 91462I, \doi{10.1117/12.2055516}

\bibitem[{{Martinache}(2016)}]{Martinache:2016}
{Martinache} F (2016) {Spectrally dispersed Fourier-phase analysis for
  redundant apertures}. In: Optical and Infrared Interferometry and Imaging V,
  \procspie, vol 9907, p 990712, \doi{10.1117/12.2233395}

\bibitem[{{Matter} et~al(2010){Matter}, {Vannier}, {Morel}, {Lopez}, {Jaffe},
  {Lagarde}, {Petrov}, and {Leinert}}]{Matter:2010}
{Matter} A, {Vannier} M, {Morel} S, {Lopez} B, {Jaffe} W, {Lagarde} S, {Petrov}
  RG, {Leinert} C (2010) {First step to detect an extrasolar planet using
  simultaneous observations with the VLTI instruments AMBER and MIDI}. \aap
  515:A69, \doi{10.1051/0004-6361/200913142}

\bibitem[{{Matter} et~al(2016{\natexlab{a}}){Matter}, {Lagarde}, {Petrov},
  {Berio}, {Robbe-Dubois}, {Lopez}, {Antonelli}, {Allouche}, {Cruzalebes},
  {Millour}, {Bazin}, and {Bourg{\`e}s}}]{Matter:2016b}
{Matter} A, {Lagarde} S, {Petrov} RG, {Berio} P, {Robbe-Dubois} S, {Lopez} B,
  {Antonelli} P, {Allouche} F, {Cruzalebes} P, {Millour} F, {Bazin} G,
  {Bourg{\`e}s} L (2016{\natexlab{a}}) {MATISSE: specifications and expected
  performances}. In: Optical and Infrared Interferometry and Imaging V,
  \procspie, vol 9907, p 990728, \doi{10.1117/12.2233542}, \eprint{1608.02351}

\bibitem[{{Matter} et~al(2016{\natexlab{b}}){Matter}, {Lopez}, {Antonelli},
  {Lehmitz}, {Bettonvil}, {Beckmann}, {Lagarde}, {Jaffe}, {Petrov}, {Berio},
  {Millour}, {Robbe-Dubois}, {Glindemann}, {Bristow}, {Schoeller}, {Lanz},
  {Henning}, {Weigelt}, {Heininger}, {Morel}, {Cruzalebes}, {Meisenheimer},
  {Hofferbert}, {Wolf}, {Bresson}, {Agocs}, {Allouche}, {Augereau}, {Avila},
  {Bailet}, {Behrend}, {van Belle}, {Berger}, {van Boekel}, {Bourget}, {Brast},
  {Clausse}, {Connot}, {Conzelmann}, {Csepany}, {Danchi}, {Delbo}, {Dominik},
  {van Duin}, {Elswijk}, {Fantei}, {Finger}, {Gabasch}, {Gont{\'e}}, {Graser},
  {Guitton}, {Guniat}, {De Haan}, {Haguenauer}, {Hanenburg}, {Hofmann},
  {Hogerheijde}, {ter Horst}, {Hron}, {Hummel}, {Isderda}, {Ives}, {Jakob},
  {Jasko}, {Jolley}, {Kiraly}, {Kragt}, {Kroener}, {Kroes}, {Kuindersma},
  {Labadie}, {Laun}, {Leinert}, {Lizon}, {Lucuix}, {Marcotto}, {Martinache},
  {Martinot-Lagarde}, {Mauclert}, {Mehrgan}, {Meilland}, {Mellein}, {Menardi},
  {Merand}, {Neumann}, {Nussbaum}, {Ottogalli}, {Palsa}, {Panduro}, {Pantin},
  {Percheron}, {Phan Duc}, {Pott}, {Pozna}, {Roelfsema}, {Rupprecht},
  {Schertl}, {Schmidt}, {Schuil}, {Spang}, {Stegmeier}, {Tromp}, {Vakili},
  {Vannier}, {Wagner}, {Venema}, and {Woillez}}]{Matter:2016a}
{Matter} A, {Lopez} B, {Antonelli} P, {Lehmitz} M, {Bettonvil} F, {Beckmann} U,
  {Lagarde} S, {Jaffe} W, {Petrov} R, {Berio} P, {Millour} F, {Robbe-Dubois} S,
  {Glindemann} A, {Bristow} P, {Schoeller} M, {Lanz} T, {Henning} T, {Weigelt}
  G, {Heininger} M, {Morel} S, {Cruzalebes} P, {Meisenheimer} K, {Hofferbert}
  R, {Wolf} S, {Bresson} Y, {Agocs} T, {Allouche} F, {Augereau} JC, {Avila} G,
  {Bailet} C, {Behrend} J, {van Belle} G, {Berger} JP, {van Boekel} R,
  {Bourget} P, {Brast} R, {Clausse} JM, {Connot} C, {Conzelmann} R, {Csepany}
  G, {Danchi} WC, {Delbo} M, {Dominik} C, {van Duin} A, {Elswijk} E, {Fantei}
  Y, {Finger} G, {Gabasch} A, {Gont{\'e}} F, {Graser} U, {Guitton} F, {Guniat}
  S, {De Haan} M, {Haguenauer} P, {Hanenburg} H, {Hofmann} KH, {Hogerheijde} M,
  {ter Horst} R, {Hron} J, {Hummel} C, {Isderda} J, {Ives} D, {Jakob} G,
  {Jasko} A, {Jolley} P, {Kiraly} S, {Kragt} J, {Kroener} T, {Kroes} G,
  {Kuindersma} S, {Labadie} L, {Laun} W, {Leinert} C, {Lizon} JL, {Lucuix} C,
  {Marcotto} A, {Martinache} F, {Martinot-Lagarde} G, {Mauclert} N, {Mehrgan}
  L, {Meilland} A, {Mellein} M, {Menardi} S, {Merand} A, {Neumann} U,
  {Nussbaum} E, {Ottogalli} S, {Palsa} R, {Panduro} J, {Pantin} E, {Percheron}
  I, {Phan Duc} T, {Pott} JU, {Pozna} E, {Roelfsema} R, {Rupprecht} G,
  {Schertl} D, {Schmidt} C, {Schuil} M, {Spang} A, {Stegmeier} J, {Tromp} N,
  {Vakili} F, {Vannier} M, {Wagner} K, {Venema} L, {Woillez} J
  (2016{\natexlab{b}}) {An overview of the mid-infrared spectro-interferometer
  MATISSE: science, concept, and current status}. In: Optical and Infrared
  Interferometry and Imaging V, \procspie, vol 9907, p 99070A,
  \doi{10.1117/12.2233052}, \eprint{1608.02350}

\bibitem[{{Meisner} and {Le Poole}(2003)}]{Meisner:2003}
{Meisner} JA, {Le Poole} RS (2003) {Dispersion affecting the VLTI and 10 micron
  interferometry using MIDI}. In: {Traub} WA (ed) Interferometry for Optical
  Astronomy II, \procspie, vol 4838, pp 609--624, \doi{10.1117/12.459072}

\bibitem[{{Mennesson} et~al(2011){Mennesson}, {Hanot}, {Serabyn}, {Liewer},
  {Martin}, and {Mawet}}]{Mennesson:2011}
{Mennesson} B, {Hanot} C, {Serabyn} E, {Liewer} K, {Martin} SR, {Mawet} D
  (2011) {High-contrast Stellar Observations within the Diffraction Limit at
  the Palomar Hale Telescope}. \apj 743:178, \doi{10.1088/0004-637X/743/2/178}

\bibitem[{{Mennesson} et~al(2013){Mennesson}, {Absil}, {Lebreton}, {Augereau},
  {Serabyn}, {Colavita}, {Millan-Gabet}, {Liu}, {Hinz}, and
  {Th{\'e}bault}}]{men13}
{Mennesson} B, {Absil} O, {Lebreton} J, {Augereau} JC, {Serabyn} E, {Colavita}
  MM, {Millan-Gabet} R, {Liu} W, {Hinz} P, {Th{\'e}bault} P (2013) {An
  Interferometric Study of the Fomalhaut Inner Debris Disk. II. Keck Nuller
  Mid-infrared Observations}. \apj 763:119, \doi{10.1088/0004-637X/763/2/119},
  \eprint{1211.7143}

\bibitem[{{Millan-Gabet} et~al(2011){Millan-Gabet}, {Serabyn}, {Mennesson},
  {Traub}, {Barry}, {Danchi}, {Kuchner}, {Stark}, {Ragland}, {Hrynevych},
  {Woillez}, {Stapelfeldt}, {Bryden}, {Colavita}, and {Booth}}]{mil11}
{Millan-Gabet} R, {Serabyn} E, {Mennesson} B, {Traub} WA, {Barry} RK, {Danchi}
  WC, {Kuchner} M, {Stark} CC, {Ragland} S, {Hrynevych} M, {Woillez} J,
  {Stapelfeldt} K, {Bryden} G, {Colavita} MM, {Booth} AJ (2011) {Exozodiacal
  Dust Levels for Nearby Main-sequence Stars: A Survey with the Keck
  Interferometer Nuller}. \apj 734:67, \doi{10.1088/0004-637X/734/1/67},
  \eprint{1104.1382}

\bibitem[{Minardi and Pertsch(2010)}]{Minardi:2010}
Minardi S, Pertsch T (2010) Interferometric beam combination with discrete
  optics. Opt Lett 35:3009--3011

\bibitem[{{Monnier}(2000)}]{Monnier:2000}
{Monnier} JD (2000) {An Introduction to Closure Phases}. In: {Lawson} PR (ed)
  Principles of Long Baseline Stellar Interferometry, p 203

\bibitem[{{Monnier} et~al(2004){Monnier}, {Berger}, {Millan-Gabet}, and {ten
  Brummelaar}}]{Monnier:2004}
{Monnier} JD, {Berger} JP, {Millan-Gabet} R, {ten Brummelaar} TA (2004) {The
  Michigan Infrared Combiner (MIRC): IR imaging with the CHARA Array}. In:
  {Traub} WA (ed) New Frontiers in Stellar Interferometry, \procspie, vol 5491,
  p 1370, \doi{10.1117/12.550804}

\bibitem[{{Monnier} et~al(2016){Monnier}, {Ireland}, {Kraus}, {Baron},
  {Creech-Eakman}, {Dong}, {Isella}, {Merand}, {Michael}, {Minardi},
  {Mozurkewich}, {Petrov}, {Rinehart}, {ten Brummelaar}, {Vasisht}, {Wishnow},
  {Young}, and {Zhu}}]{Monnier:2016}
{Monnier} JD, {Ireland} MJ, {Kraus} S, {Baron} F, {Creech-Eakman} M, {Dong} R,
  {Isella} A, {Merand} A, {Michael} E, {Minardi} S, {Mozurkewich} D, {Petrov}
  R, {Rinehart} S, {ten Brummelaar} T, {Vasisht} G, {Wishnow} E, {Young} J,
  {Zhu} Z (2016) {Architecture design study and technology road map for the
  Planet Formation Imager (PFI)}. In: Optical and Infrared Interferometry and
  Imaging V, \procspie, vol 9907, p 99071O, \doi{10.1117/12.2233311},
  \eprint{1608.00580}

\bibitem[{{Mordasini} et~al(2012){Mordasini}, {Alibert}, {Klahr}, and
  {Henning}}]{Mordasini:2012}
{Mordasini} C, {Alibert} Y, {Klahr} H, {Henning} T (2012) {Characterization of
  exoplanets from their formation. I. Models of combined planet formation and
  evolution}. \aap 547:A111, \doi{10.1051/0004-6361/201118457},
  \eprint{1206.6103}

\bibitem[{Nguyen et~al(2017)Nguyen, R\'{o}denas, de~Aldana, Mart\'{i}n,
  Mart\'{i}nez, Aguil\'{o}, Pujol, and D\'{i}az}]{Nguyen:2017}
Nguyen HD, R\'{o}denas A, de~Aldana JRV, Mart\'{i}n G, Mart\'{i}nez J,
  Aguil\'{o} M, Pujol MC, D\'{i}az F (2017) Low-loss 3d-laser-written
  mid-infrared linbo3 depressed-index cladding waveguides for both te and tm
  polarizations. Opt Express 25(4):3722--3736, \doi{10.1364/OE.25.003722},
  \urlprefix\url{http://www.opticsexpress.org/abstract.cfm?URI=oe-25-4-3722}

\bibitem[{{Norris} et~al(2014){Norris}, {Cvetojevic}, {Gross}, {Jovanovic},
  {Stewart}, {Charles}, {Lawrence}, {Withford}, and {Tuthill}}]{Norris:2014}
{Norris} B, {Cvetojevic} N, {Gross} S, {Jovanovic} N, {Stewart} PN, {Charles}
  N, {Lawrence} JS, {Withford} MJ, {Tuthill} P (2014) {High-performance 3D
  waveguide architecture for astronomical pupil-remapping interferometry}.
  Optics Express 22:18,335, \doi{10.1364/OE.22.018335}, \eprint{1405.7428}

\bibitem[{{Olofsson} et~al(2013){Olofsson}, {Benisty}, {Le Bouquin}, {Berger},
  {Lacour}, {M{\'e}nard}, {Henning}, {Crida}, {Burtscher}, {Meeus}, {Ratzka},
  {Pinte}, {Augereau}, {Malbet}, {Lazareff}, and {Traub}}]{Olofsson:2013}
{Olofsson} J, {Benisty} M, {Le Bouquin} JB, {Berger} JP, {Lacour} S,
  {M{\'e}nard} F, {Henning} T, {Crida} A, {Burtscher} L, {Meeus} G, {Ratzka} T,
  {Pinte} C, {Augereau} JC, {Malbet} F, {Lazareff} B, {Traub} W (2013)
  {Sculpting the disk around T Chamaeleontis: an interferometric view}. \aap
  552:A4, \doi{10.1051/0004-6361/201220675}, \eprint{1302.2622}

\bibitem[{{Pott} et~al(2012){Pott}, {M{\"u}ller}, {Karovicova}, and
  {Delplancke}}]{Pott:2012}
{Pott} JU, {M{\"u}ller} A, {Karovicova} I, {Delplancke} F (2012) {New horizons
  for VLTI 10 micron interferometry: first scientific measurements with
  external PRIMA fringe tracking}. In: Optical and Infrared Interferometry III,
  \procspie, vol 8445, p 84450Q, \doi{10.1117/12.927027}

\bibitem[{{Pott} et~al(2016){Pott}, {Fu}, {Widmann}, and {Peter}}]{Pott:2016}
{Pott} JU, {Fu} Q, {Widmann} F, {Peter} D (2016) {P-REx: the piston drift
  reconstruction experiment}. In: Optical and Infrared Interferometry and
  Imaging V, \procspie, vol 9907, p 99073E, \doi{10.1117/12.2233139}

\bibitem[{Rodenas et~al(2012)Rodenas, Martin, Arzeki, Psaila, Jose, Jha,
  Labadie, Kern, Kar, and Thomson}]{Rodenas:2012}
Rodenas A, Martin G, Arzeki B, Psaila ND, Jose G, Jha A, Labadie L, Kern P, Kar
  AK, Thomson RR (2012) Three-dimensional mid-infrared photonic circuits in
  chalcogenide glass. Opt Lett 37:392--394

\bibitem[{{Roettenbacher} et~al(2015){Roettenbacher}, {Monnier}, {Fekel},
  {Henry}, {Korhonen}, {Latham}, {Muterspaugh}, {Williamson}, {Baron}, {ten
  Brummelaar}, {Che}, {Harmon}, {Schaefer}, {Scott}, {Sturmann}, {Sturmann},
  and {Turner}}]{Roettenbacher:2015}
{Roettenbacher} RM, {Monnier} JD, {Fekel} FC, {Henry} GW, {Korhonen} H,
  {Latham} DW, {Muterspaugh} MW, {Williamson} MH, {Baron} F, {ten Brummelaar}
  TA, {Che} X, {Harmon} RO, {Schaefer} GH, {Scott} NJ, {Sturmann} J, {Sturmann}
  L, {Turner} NH (2015) {Detecting the Companions and Ellipsoidal Variations of
  RS CVn Primaries. II. o Draconis, a Candidate for Recent Low-mass Companion
  Ingestion}. \apj 809:159, \doi{10.1088/0004-637X/809/2/159},
  \eprint{1507.03601}

\bibitem[{{Sana} et~al(2014){Sana}, {Le Bouquin}, {Lacour}, {Berger}, {Duvert},
  {Gauchet}, {Norris}, {Olofsson}, {Pickel}, {Zins}, {Absil}, {de Koter},
  {Kratter}, {Schnurr}, and {Zinnecker}}]{Sana:2014}
{Sana} H, {Le Bouquin} JB, {Lacour} S, {Berger} JP, {Duvert} G, {Gauchet} L,
  {Norris} B, {Olofsson} J, {Pickel} D, {Zins} G, {Absil} O, {de Koter} A,
  {Kratter} K, {Schnurr} O, {Zinnecker} H (2014) {Southern Massive Stars at
  High Angular Resolution: Observational Campaign and Companion Detection}.
  \apjs 215:15, \doi{10.1088/0067-0049/215/1/15}, \eprint{1409.6304}

\bibitem[{{Serabyn}(2000)}]{Serabyn:2000}
{Serabyn} E (2000) {Nulling interferometry: symmetry requirements and
  experimental results}. In: {L{\'e}na} P, {Quirrenbach} A (eds) Interferometry
  in Optical Astronomy, \procspie, vol 4006, pp 328--339,
  \doi{10.1117/12.390223}

\bibitem[{{Snellen} et~al(2015){Snellen}, {de Kok}, {Birkby}, {Brandl},
  {Brogi}, {Keller}, {Kenworthy}, {Schwarz}, and {Stuik}}]{Snellen:2015}
{Snellen} I, {de Kok} R, {Birkby} JL, {Brandl} B, {Brogi} M, {Keller} C,
  {Kenworthy} M, {Schwarz} H, {Stuik} R (2015) {Combining high-dispersion
  spectroscopy with high contrast imaging: Probing rocky planets around our
  nearest neighbors}. \aap 576:A59, \doi{10.1051/0004-6361/201425018},
  \eprint{1503.01136}

\bibitem[{{Spiegel} and {Burrows}(2012)}]{Spiegel:2012}
{Spiegel} DS, {Burrows} A (2012) {Spectral and Photometric Diagnostics of Giant
  Planet Formation Scenarios}. \apj 745:174, \doi{10.1088/0004-637X/745/2/174},
  \eprint{1108.5172}

\bibitem[{{Tepper} et~al(2017){Tepper}, {Labadie}, {Gross}, {Arriola},
  {Minardi}, {Diener}, and {Withford}}]{Tepper:2017b}
{Tepper} J, {Labadie} L, {Gross} S, {Arriola} A, {Minardi} S, {Diener} R,
  {Withford} MJ (2017) {Ultrafast laser inscription in ZBLAN integrated optics
  chips for mid-IR beam combination in astronomical interferometry}. Optics
  Express 25:20,642, \doi{10.1364/OE.25.020642}

\bibitem[{{Tepper, J.} et~al(2017){Tepper, J.}, {Labadie, L.}, {Diener, R.},
  {Minardi, S.}, {Pott, J.-U.}, {Thomson, R.}, and {Nolte, S.}}]{Tepper:2017a}
{Tepper, J}, {Labadie, L}, {Diener, R}, {Minardi, S}, {Pott, J-U}, {Thomson,
  R}, {Nolte, S} (2017) Integrated optics prototype beam combiner for long
  baseline interferometry in the l and m bands. A\&A 602:A66,
  \doi{10.1051/0004-6361/201630138},
  \urlprefix\url{https://doi.org/10.1051/0004-6361/201630138}

\bibitem[{{Tristram} and {Schartmann}(2011)}]{Tristram:2011}
{Tristram} KRW, {Schartmann} M (2011) {On the size-luminosity relation of AGN
  dust tori in the mid-infrared}. \aap 531:A99,
  \doi{10.1051/0004-6361/201116867}, \eprint{1105.4875}

\bibitem[{{Tristram} et~al(2014){Tristram}, {Burtscher}, {Jaffe},
  {Meisenheimer}, {H{\"o}nig}, {Kishimoto}, {Schartmann}, and
  {Weigelt}}]{Tristram:2014}
{Tristram} KRW, {Burtscher} L, {Jaffe} W, {Meisenheimer} K, {H{\"o}nig} SF,
  {Kishimoto} M, {Schartmann} M, {Weigelt} G (2014) {The dusty torus in the
  Circinus galaxy: a dense disk and the torus funnel}. \aap 563:A82,
  \doi{10.1051/0004-6361/201322698}, \eprint{1312.4534}

\bibitem[{{Willson} et~al(2016){Willson}, {Kraus}, {Kluska}, {Monnier},
  {Ireland}, {Aarnio}, {Sitko}, {Calvet}, {Espaillat}, and
  {Wilner}}]{Willson:2016}
{Willson} M, {Kraus} S, {Kluska} J, {Monnier} JD, {Ireland} M, {Aarnio} A,
  {Sitko} ML, {Calvet} N, {Espaillat} C, {Wilner} DJ (2016) {Sparse aperture
  masking interferometry survey of transitional discs. Search for
  substellar-mass companions and asymmetries in their parent discs}. \aap
  595:A9, \doi{10.1051/0004-6361/201628859}, \eprint{1608.03629}

\end{thebibliography}

\end{document}